\documentclass[final,5p,times,twocolumn]{elsarticle}

\usepackage{amssymb}
\usepackage{amsmath}

\usepackage{comment}
\usepackage{xurl}
\usepackage{amsmath,amssymb,amsfonts}
\usepackage{algorithmic}
\usepackage{graphicx}
\usepackage{textcomp}
\usepackage[dvipsnames]{xcolor}
\usepackage{comment}
\usepackage{array}
\usepackage{multirow}
\usepackage{caption}
\usepackage{subcaption}
\usepackage{tikz}
\usepackage{adjustbox}
\usepackage{soul}
\usepackage{booktabs}
\usepackage{tabularx}
\usepackage{bbding}
\usepackage{pifont}
\usepackage{wasysym}
\usepackage{amssymb}
\def\BibTeX{{\rm B\kern-.05em{\sc i\kern-.025em b}\kern-.08em
    T\kern-.1667em\lower.7ex\hbox{E}\kern-.125emX}}
\definecolor{green}{rgb}{0.09, 0.45, 0.27}
\usepackage{comment}
\usepackage{setspace}
\usepackage{todonotes}
\usepackage{pdfcomment}
\usepackage{fancyhdr}

\newlength\MAX  

\newcommand{\DrawPercentageBar}[1]{%
  \begin{tikzpicture}
    \fill[color=black]   (0.0 , 0.0) rectangle (#1*3ex , 1.5ex );
    \fill[color=gray] (#1*3ex  , 0.0) rectangle (3.0ex, 1.5ex);
  \end{tikzpicture}%
}
\newcommand{\DrawPercentageBarRed}[1]{%
  \begin{tikzpicture}
    \fill[color=red]   (0.0 , 0.0) rectangle (#1*3ex , 1.5ex );
    \fill[color=gray] (#1*3ex  , 0.0) rectangle (3.0ex, 1.5ex);
  \end{tikzpicture}%
}
\newcommand{\DrawPercentageBarBlue}[1]{%
  \begin{tikzpicture}
    \fill[color=blue]   (0.0 , 0.0) rectangle (#1*3ex , 1.5ex );
    \fill[color=gray] (#1*3ex  , 0.0) rectangle (3.0ex, 1.5ex);
  \end{tikzpicture}%
}
\newcommand*\circled[1]{\tikz[baseline=(char.base)]{
            \node[shape=circle,draw,inner sep=1pt,font=\sffamily\footnotesize] (char) {\textbf{#1}};}}

\usepackage[most]{tcolorbox} 
 
\definecolor{lightgreen}{rgb}{0.894, 0.961, 0.949}
\definecolor{darkgreen}{rgb}{0.0, 0.576, 0.533}
\definecolor{lightblue}{rgb}{0.88, 0.95, 1.0} 
\definecolor{darkblue}{rgb}{0.0, 0.2, 0.6} 
 
\tcbset {
  base/.style={
    enhanced,
    breakable,
    arc=0mm, 
    boxrule=0mm,
    colback=lightblue!20!, 
    left=3.5mm,
    leftrule=2mm, 
    right=3.5mm,
  }
}
 
\newtcolorbox{mainbox}[1]{
  colframe=darkblue, 
  base={#1}
}

\journal{}

\begin{document}

\begin{frontmatter}



\title{Elevating Cyber Threat Intelligence against Disinformation Campaigns\\with LLM-based Concept Extraction and the FakeCTI Dataset} 


\author[dieti]{Domenico Cotroneo} 
\author[gssi]{Roberto Natella} 
\author[dieti]{Vittorio Orbinato} 

\affiliation[dieti]{organization={DIETI, Università degli Studi di Napoli Federico II},
            addressline={via Claudio 21}, 
            city={Naples},
            postcode={80125}, 
            country={Italy}}
\affiliation[gssi]{organization={Gran Sasso Science Institute},
            city={L'Aquila},
            country={Italy}}

\begin{abstract}  
The swift spread of fake news and disinformation campaigns poses a significant threat to public trust, political stability, and cybersecurity. Traditional Cyber Threat Intelligence (CTI) approaches, which rely on low-level indicators such as domain names and social media handles, are easily evaded by adversaries who frequently modify their online infrastructure. To address these limitations, we introduce a novel CTI framework that focuses on high-level, semantic indicators derived from recurrent narratives and relationships of disinformation campaigns. 
Our approach extracts structured CTI indicators from unstructured disinformation content, capturing key entities and their contextual dependencies within fake news using Large Language Models (LLMs). We further introduce FakeCTI, the first dataset that systematically links fake news to disinformation campaigns and threat actors. To evaluate the effectiveness of our CTI framework, we analyze multiple fake news attribution techniques, spanning from traditional Natural Language Processing (NLP) to fine-tuned LLMs. 
This work shifts the focus from low-level artifacts to persistent conceptual structures, establishing a scalable and adaptive approach to tracking and countering disinformation campaigns. 
\end{abstract}




\end{frontmatter}


\section{Introduction}
\label{sec:introduction}
In recent years, the proliferation of fake news has emerged as a critical threat to global information integrity. Fabricated with misleading and false content, fake news is designed to misinform, manipulate, and deceive audiences to serve specific agendas \cite{shu2019studying, meel2020fake, PoliticoAI}. Malicious actors craft these narratives to influence public discourse, sway political decisions, and disrupt societal harmony. The widespread adoption of social media platforms has further aggravated this issue, enabling rapid dissemination of disinformation at an unprecedented scale \cite{EuroFakeNews, UnescoFakeNews, aimeur2023fake, tacchini2017some}. As a result, distinguishing between authentic and deceptive content has become increasingly challenging.

To combat this growing threat, security analysts leverage \textit{Cyber Threat Intelligence} (CTI) to track and mitigate disinformation campaigns. This CTI includes links to web articles and social media posts classified by analysts as fake content, along with the domains and social media accounts that created them. However, this kind of CTI is affected by the same issues as the ``traditional'' CTI used for network attacks, which includes web domains, IP addresses, and hostnames. Attackers can easily modify their attacks to make them unrecognizable by CTI, e.g., by buying new domains. The same logic applies to disinformation and fake news, where attackers can create new websites or social media accounts to evade detection. 

\begin{figure*}[htbp]
    \centering
    \includegraphics[width=0.75\textwidth]{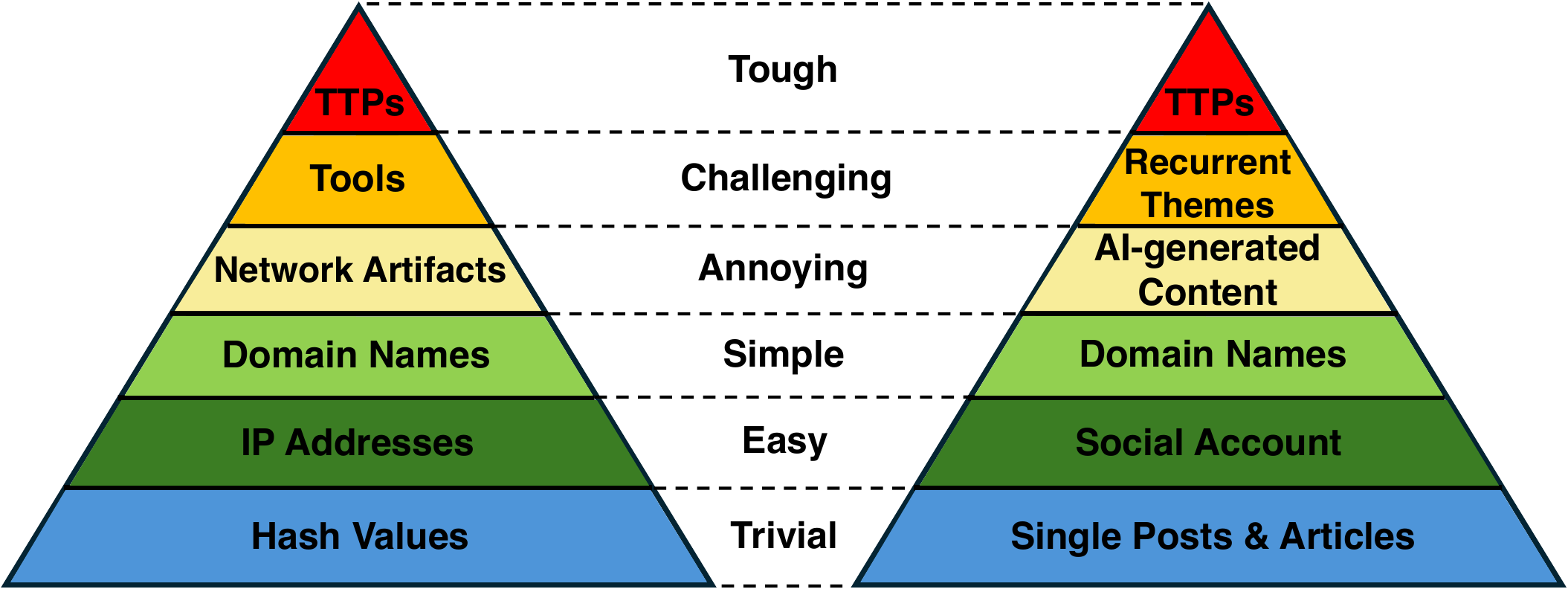}
    \caption{Traditional and disinformation-oriented Pyramids of Pain.}
    \label{fig:Pyramids}
\end{figure*}

This issue is highlighted by the \textit{Pyramid of Pain} \citep{pyramidofpain}, a framework designed for network-based CTI that classifies low-level indicators as trivial to bypass, while higher-level intelligence is significantly harder for adversaries to evade. Figure \ref{fig:Pyramids} depicts how we adapt the Pyramid of Pain to the disinformation landscape, illustrating how different types of intelligence vary in their persistence and resistance to manipulation. To be effective, CTI for fake news must move beyond easily modifiable artifacts and capture persistent elements of disinformation operations. Our adaptation of the Pyramid of Pain for disinformation shifts the focus from technical artifacts to content-based indicators. While traditional CTI relies on low-level indicators such as IP addresses, domains, and hashes, which adversaries can rapidly modify, disinformation campaigns similarly exploit website domains and social media handles that are easily replaced. To develop a more resilient strategy, our model shifts the focus to persistent narrative structures and behavioral patterns that support disinformation operations.
Unlike network artifacts, core narratives cannot be easily rewritten, as they structure the fundamental viewpoint promoted by threat actors. 

This work aims to extract the fundamental elements that characterize a fake news campaign, in order to recognize new fake content from the same attacker in a way that is robust to textual modifications and variations of domains and accounts.
To address the limitations of traditional CTI approaches in tackling disinformation, we introduce a concept-based CTI framework that captures the narratives and relationships within disinformation campaigns. Unlike conventional CTI indicators, which primarily rely on low-level technical artifacts, our methodology extracts high-level, content-based intelligence that persists even as threat actors modify superficial elements of their campaigns. Our approach is based on the structured representation of disinformation narratives through tuples that encapsulate the key entities, their relationships, and contextual dependencies within fake news articles. These structured representations allow for more robust attribution and variant detection, as they focus on the core meaning and intent rather than specific linguistic expressions. We employ Large Language Models (LLMs) to automate the extraction and analysis of these structured indicators, effectively converting unstructured narratives into structured intelligence. In addition, we analyze multiple analytical techniques, ranging from lexical and semantic similarity methods to fine-tuned LLMs, to recognize new variants of fake news that align with existing disinformation campaigns, even when the language and presentation of the content have been altered.

The primary contributions of this study are as follows:
\begin{enumerate}
\item \textit{Concept-based CTI indicators extraction}: we propose a novel methodology to extract structured CTI indicators that characterize fake news narratives through key entities, relationships, and their contextual connections. This structured representation enables robust and adaptable attribution by focusing on thematic and relational consistency. By encoding the core meaning of fake news content, these indicators enable the detection of new campaign variants and the association of evolving disinformation efforts with previously identified attack groups. The extracted intelligence is represented as structured tuples that capture the essential actions and relationships in disinformation narratives, ensuring a scalable and consistent format for further analysis.
\item \textit{FakeCTI dataset}: to facilitate research in this domain, we introduce FakeCTI, the first dataset that systematically links fake news articles to known disinformation campaigns and threat actors. FakeCTI includes $12,155$ articles from $43$ distinct campaigns, each annotated with metadata specifying the associated campaign, threat actor, and dissemination medium. FakeCTI allows researchers and practitioners to analyze disinformation trends, develop automated detection systems, and evaluate attribution methodologies at scale, offering a structured foundation for attribution studies.
\item \textit{Experimental evaluation of fake news attribution techniques}: using the FakeCTI dataset, we assess multiple techniques for attributing fake news articles to their originating disinformation campaigns. Our evaluation spans several approaches, achieving accuracy in fake news attribution up to 94\% (using LLMs), highlighting how well different techniques generalize across disinformation content and their capability to identify new and evolving fake news narratives.
\end{enumerate}

This study demonstrates how concept-based CTI indicators can help improving disinformation detection, attribution, and prevention, offering a more effective approach to countering manipulated narratives and adversarial disinformation strategies. We publicly share our artifacts\footnote{\href{https://github.com/dessertlab/Concept-based-Disinformation-CTI}{Code: https://github.com/dessertlab/Concept-based-Disinformation-CTI}}\footnote{\href{https://zenodo.org/records/17248748}{FakeCTI Dataset: https://zenodo.org/records/17248748}} to encourage further research on this topic.

The remainder of the paper is structured as follows. Section \ref{sec:related} discusses related work; Section \ref{sec:methodology} presents the methodology for the extraction of concept-based CTI indicators; Section \ref{sec:dataset} introduces the FakeCTI dataset; Section \ref{sec:questions}, \ref{sec:experiment} illustrate the experimental analysis on the indicators extraction and fake news attribution; Section \ref{sec:conclusion} concludes the paper.

\section{Background and Related Work}
\label{sec:related}
\subsection{Fake News Detection}
Sharma \emph{et al.} \citep{sharma2019combating} performed a comprehensive analysis of fake news detection and mitigation techniques, categorizing existing methodologies into three main approaches: content-based identification, feedback-based identification, and intervention-based solutions.

\vspace{3pt}
\noindent
\textbf{Content-based identification}. Content-based fake news detection techniques analyze the text and linguistic features of news articles to extract distinguishing characteristics between fake and true news and identify deceptive patterns \citep{mahyoob2020linguistic, choudhary2021linguistic, pan2018content}. These methods include linguistic-based techniques, which try to identify deception through stylistic markers and factual inconsistencies, and deep learning-based solutions, which aim to extract complex textual patterns without manual feature engineering. However, adversaries can evade this kind of detection by rewording or slightly modifying content. Simple paraphrasing or reformatting headlines can bypass content-based classifiers, making such models ineffective against large-scale disinformation campaigns.


\vspace{3pt}
\noindent
\textbf{Feedback-based identification}. Feedback-based methods leverage user interactions, propagation patterns, and engagement metrics on social media to detect fake news. These strategies include propagation pattern, temporal pattern, and user response and stance analysis, which assume that the way information spreads can indicate its veracity \citep{liu2018early, monti2019fake, ruchansky2017csi}. A significant limitation of many existing approaches is their focus on single social media platforms such as Twitter/X or Facebook. Consequently, adversaries easily evade detection by shifting narratives across different ecosystems, such as migrating from one platform to another, e.g., from Twitter/X to Telegram or Reddit, when content moderation policies increase. Strengthening detection requires integrating multi-platform tracking systems that correlate narratives, domain registrations, and user interactions across different platforms.


\vspace{3pt}
\noindent
\textbf{Intervention-based solutions}. Intervention-based solutions represent mitigation strategies to prevent the spread of disinformation. These approaches include proactive disinformation interventions, such as fact-checking bots, which analyze claims against verified knowledge bases to disprove false information, credibility scores assigned to sources, content moderation, consisting of machine learning-based techniques to reduce the reach of fake news by limiting algorithmic amplification, and user awareness and debunking strategies \citep{bak2022combining, ciampaglia2015computational, popat2018declare}. However, fact-checking and debunking approaches are typically slow, while content moderation strategies must balance accuracy with freedom of speech concerns.

As highlighted above, most existing detection techniques focus on identifiable artifacts such as suspicious domain registrations, automated social media accounts, or content propagation anomalies. While these features provide valuable clues about the spread of disinformation, they remain reactive indicators that adversaries can quickly replace or modify. 
Indeed, adversaries can easily generate new social media accounts or register new domains once previous ones are flagged. To effectively disrupt disinformation, cybersecurity teams must analyze the tactics, techniques, and procedures (TTPs) used by adversaries, such as narrative amplification techniques, AI-generated content, and cross-platform coordination tactics.

Another significant shortcoming is the lack of attribution and integration with traditional CTI workflows. While frameworks like STIX \citep{stix, STIXMITRE} and TAXII \citep{TAXIIMITRE} enable structured CTI sharing for cyber threats, disinformation analysis remains fragmented. Current approaches, such as fact-checking databases and disinformation-debunking sites, operate in isolation from security operations centers (SOCs) and intelligence-sharing platforms, e.g., MISP \citep{MISP}. Detection and attribution efforts should focus on the highest levels of the Pyramid of Pain, \textit{Recurrent Themes} and \textit{TTPs}, where adversaries rely on known strategies but change execution methods. To strengthen attribution, CTI workflows must incorporate standardized frameworks, such as DISARM, that map disinformation actors and TTPs into structured intelligence feeds. By aligning disinformation intelligence with standardized cybersecurity methodologies, organizations can develop multi-layered defense mechanisms, allowing for real-time intelligence-sharing across governments, tech platforms, and fact-checking networks.

\subsection{Disinformation-oriented CTI}
In addition to traditional fake news identification and mitigation techniques, the \textit{DISARM} (Disinformation Analysis and Response Measures) framework \citep{DISARM} has been proposed as a standardized approach for integrating disinformation-oriented CTI. DISARM provides a structured approach to analyzing and countering online disinformation by categorizing threat actors, attack techniques, and mitigation strategies similar to established CTI frameworks like MITRE ATT\&CK \citep{attack}. The framework aims to enhance the interoperability of threat intelligence across organizations, social media platforms, and cybersecurity agencies by defining a common taxonomy for disinformation campaigns. Key elements of DISARM include actor profiling (e.g., state-sponsored campaigns, cybercriminals, ideological groups), tactics (e.g., narrative amplification, deepfake propagation, social engineering), and countermeasures (e.g., AI-driven detection, cross-platform attribution, content moderation). Unlike traditional fake news detection, which often focuses on individual articles or social media posts, DISARM treats disinformation as a systemic cyber threat requiring multi-layered intelligence sharing and automated threat hunting. However, challenges remain in operationalizing the framework, particularly in real-time attribution of adversarial actors and alignment of disinformation CTI with existing cybersecurity workflows.

To this end, Gonzalez \emph{et al.} \citep{gonzalez2025toward} proposed \textit{DISINFOX}, an open-source framework to integrate disinformation intelligence into existing CTI workflows, leveraging DISARM to model disinformation TTPs and translate them into STIX objects for structured representation and interoperability. DISINFOX facilitates the storage, management, and exchange of disinformation incidents, providing integration with OpenCTI \citep{OpenCTI}. The framework was validated using a dataset of 118 real-world disinformation incidents, demonstrating its feasibility for structured intelligence-sharing and analysis. However, challenges such as manual annotation, limited automation, and lack of standardized evaluation frameworks may hinder its large-scale operationalization.

Despite its structured approach, DISARM is not yet ready for practical deployment due to critical limitations. Limited adoption and institutional backing prevent its widespread integration into cybersecurity workflows, as most CTI platforms are not yet configured to handle disinformation intelligence. Moreover, DISARM lacks standardized evaluation frameworks to assess the impact and severity of disinformation incidents, unlike traditional CTI which benefits from well-established risk-scoring models, e.g., CVSS for vulnerabilities \citep{CVSS}. Determining the virality, influence, and effectiveness of a disinformation campaign is highly context-dependent, making it difficult to assign threat levels unequivocally. 

Our solution, focused on extracting and analyzing structured concept-based CTI, aligns with the \textit{Recurrent Themes} level in the disinformation-oriented Pyramid of Pain, the highest abstraction layer just below TTPs. This reflects the persistent nature of themes and narratives across campaign variants, central to attribution and detection. While TTPs remain highly impactful, their standardization is still limited, as highlighted by the challenges faced in operationalizing DISARM.

\subsection{Fake News Datasets}
The landscape of fake news analysis has evolved significantly with the availability of structured datasets designed for training machine learning models for disinformation detection. While several datasets provide valuable insights into fake news detection, their applicability to fake news attribution, especially concerning disinformation campaigns or threat actors, remains limited. 

\emph{Fakeddit} \citep{nakamura2020fakeddit} is a large-scale, multimodal dataset containing over 1 million samples collected from Reddit, a social media platform where fake news, conspiracy theories, and propaganda often circulate. It is one of the most extensive publicly available datasets for fake news detection, providing textual and visual features alongside metadata and user engagement data. Unlike many datasets that offer binary classifications (true/false), Fakeddit employs a more granular labeling system, allowing researchers to categorize fake news into six different classes: true, mostly true, half true, half false, mostly false, and false.
Despite its comprehensiveness, the dataset does not explicitly connect disinformation to specific threat actors or coordinated disinformation campaigns, nor does it provide insights into the intent behind the disinformation.

\emph{IFND} \citep{sharma2023ifnd} is a multimodal dataset that focuses on Indian fake news, covering news articles spanning from 2013 to 2021. The dataset contains both true and false news, allowing researchers to explore the characteristics of disinformation within an Indian context.
A distinctive feature of IFND is its use of Latent Dirichlet Allocation (LDA) topic modeling, which helps classify news articles into different topical categories. 
However, while the dataset enables fine-grained classification and content-based fake news detection, it does not provide any attribution to threat actors or coordinated influence campaigns. The dataset lacks metadata that could link fake news articles to disinformation networks.

The \emph{LIAR} dataset \citep{wang2017liar} is one of the most widely used benchmark datasets for fake news detection and fact-checking. It consists of 12,836 manually labeled short statements collected from Politifact \cite{Politifact}, a Pulitzer Prize-winning fact-checking organization. Unlike Fakeddit and IFND, LIAR is specifically designed to analyze political disinformation: it includes statements made by politicians, government officials, and public figures, along with detailed justifications for each truthfulness label.
Each statement in LIAR is classified into six fine-grained categories: pants-on-fire (completely false), false, barely true, half-true, mostly true, and true. While LIAR provides a structured approach to analyzing political disinformation, it does not capture broader disinformation campaigns or any organized influence operations.

The \emph{PushShift Telegram} dataset \citep{baumgartner2020pushshifttelegram} is unique since it focuses on disinformation spread through Telegram, a messaging platform known for hosting extremist groups, conspiracy theorists, and disinformation networks. Unlike previous datasets, which collect news articles and political statements, PushShift Telegram comprises messages from 27,800+ Telegram channels, capturing real-time discussions, propaganda, and conspiracy-driven content.
This dataset is particularly relevant for studying coordinated disinformation efforts, as many Telegram groups serve as hubs for extremist and disinformation networks, providing valuable insight into how disinformation spreads within closed communities.
Although it includes data from channels linked to disinformation campaigns, PushShift Telegram does not provide direct attribution to specific threat actors. Without metadata linking the groups to nation-state actors, troll farms, or intelligence operations, it remains challenging to draw definitive conclusions about the origins and intent behind the disinformation.

The \emph{PushShift Reddit} dataset \citep{baumgartner2020pushshiftreddit}, like the Telegram version, provides a large-scale collection of Reddit discussions and posts. The dataset contains millions of comments and posts from Reddit users. Unlike Fakeddit, which focuses on structured fake news classification, PushShift Reddit captures unfiltered, real-time conversations, making it valuable for studying disinformation narratives and how they evolve.
While the dataset includes discussions on conspiracies, political propaganda, and hoaxes, it does not explicitly label fake and true news.
Similar to PushShift Telegram, the dataset is useful for studying how disinformation spreads, but it does not provide direct links between disinformation campaigns and specific threat actors.

\emph{BuzzFace} \citep{santia2018buzzface} is a comprehensive and large-scale dataset designed to study fake news dissemination and user engagement on Facebook. It includes 2,282 annotated news articles spanning mainstream, left-leaning, and right-leaning news sources, providing a diverse representation of disinformation. The dataset categorizes news into four veracity levels: mostly true, mostly false, a mixture of true and false, and no factual content.
BuzzFace also includes metadata such as engagement metrics, user comments, and reactions. 
Although the dataset captures patterns of interaction and possible social bot activity, it lacks detailed attribution metadata, such as information on who created or amplified the disinformation. 

\begin{table*}[htbp]
    \caption{Comparison of existing fake news datasets.}
    \centering
    \renewcommand{\arraystretch}{1.3}
    \resizebox{\linewidth}{!}{%
    \begin{tabular}{lcccc}
        \toprule
        \textbf{Dataset} & \textbf{Samples} & \textbf{Source} & \textbf{Threat Actor Association} & \textbf{Usability for Attribution} \\
        \midrule
        Fakeddit & 1,063,106 & Reddit & No & Low \\ 
        \midrule
        IFND & Unspecified & Indian news websites & No & Low \\ 
        \midrule
        LIAR & 12,836 & Politifact & No & Low \\ 
        \midrule
        PushShift Telegram & Unspecified & Telegram & Yes & Moderate \\ 
        \midrule
        PushShift Reddit & Unspecified & Reddit & Possible & Moderate \\ 
        \midrule
        BuzzFace & 2,282 articles, 1.6M comments & Facebook & No & Low \\ 
        \midrule
        \midrule
        \textbf{FakeCTI} & 12,155 articles, 43 disinformation campaigns & News websites, Social networks & Yes & High \\
        \bottomrule
    \end{tabular}%
    }
    \label{tab:fake_news_datasets_comparison}
\end{table*}

Table \ref{tab:fake_news_datasets_comparison} summarizes the comparison between these existing datasets. It is possible to notice how most datasets lack explicit connections to threat actors or coordinated disinformation campaigns, making them unsuitable for attribution. Fakeddit and IFND, despite their large-scale nature, focus on fake news classification rather than tracing its origin, while LIAR is limited to fact-checked political statements, offering no insights into organized influence operations. BuzzFace, which integrates Facebook user engagement, provides valuable data on misinformation spread but lacks source tracking or disinformation network links. Among them, PushShift Telegram is the most relevant for threat actor association, as it captures propaganda narratives and manipulated content from extremist groups, making it a potential resource for analyzing coordinated disinformation efforts. However, it still requires external intelligence sources to attribute fake news to specific actors definitively. PushShift Reddit, though capturing discussions where misinformation evolves, does not explicitly tag or track coordinated campaigns, making its association with threat actors speculative. In terms of usability for fake news attribution, only PushShift Telegram and, to a lesser extent, PushShift Reddit, hold moderate potential. On the contrary, Fakeddit, IFND, LIAR, and BuzzFace are better suited for fake news detection and classification rather than identifying who creates or disseminates misinformation. While these datasets provide valuable insights into how misinformation spreads, they lack critical metadata for direct attribution, such as source credibility, bot network activity, or links to known disinformation operations. PushShift Telegram remains the most promising but requires additional forensic analysis and intelligence sources to confirm attribution. Effective fake news attribution demands more than just classification, it requires datasets that track source origins, coordinated amplification, and intent, none of which are fully addressed by these datasets.

Conversely, the proposed dataset, i.e., \emph{FakeCTI}, offers a significant advantage over existing fake news datasets by systematically linking fake news articles to their corresponding disinformation campaigns and threat actors. While prior datasets primarily focus on binary or multi-class fake news classification, labeling articles as true, false, or misleading, they lack explicit connections to the broader disinformation strategies that spread these narratives.
FakeCTI provides structured attribution metadata, allowing researchers to trace the origin and intent of disinformation campaigns. This linkage is crucial for effective fake news attribution, as it allows the detection of new campaign variants and the identification of recurring disinformation strategies across multiple sources. 


\section{Methodology}
\label{sec:methodology}
\begin{figure*}[ht]
\centering
\includegraphics[width=0.8\linewidth]{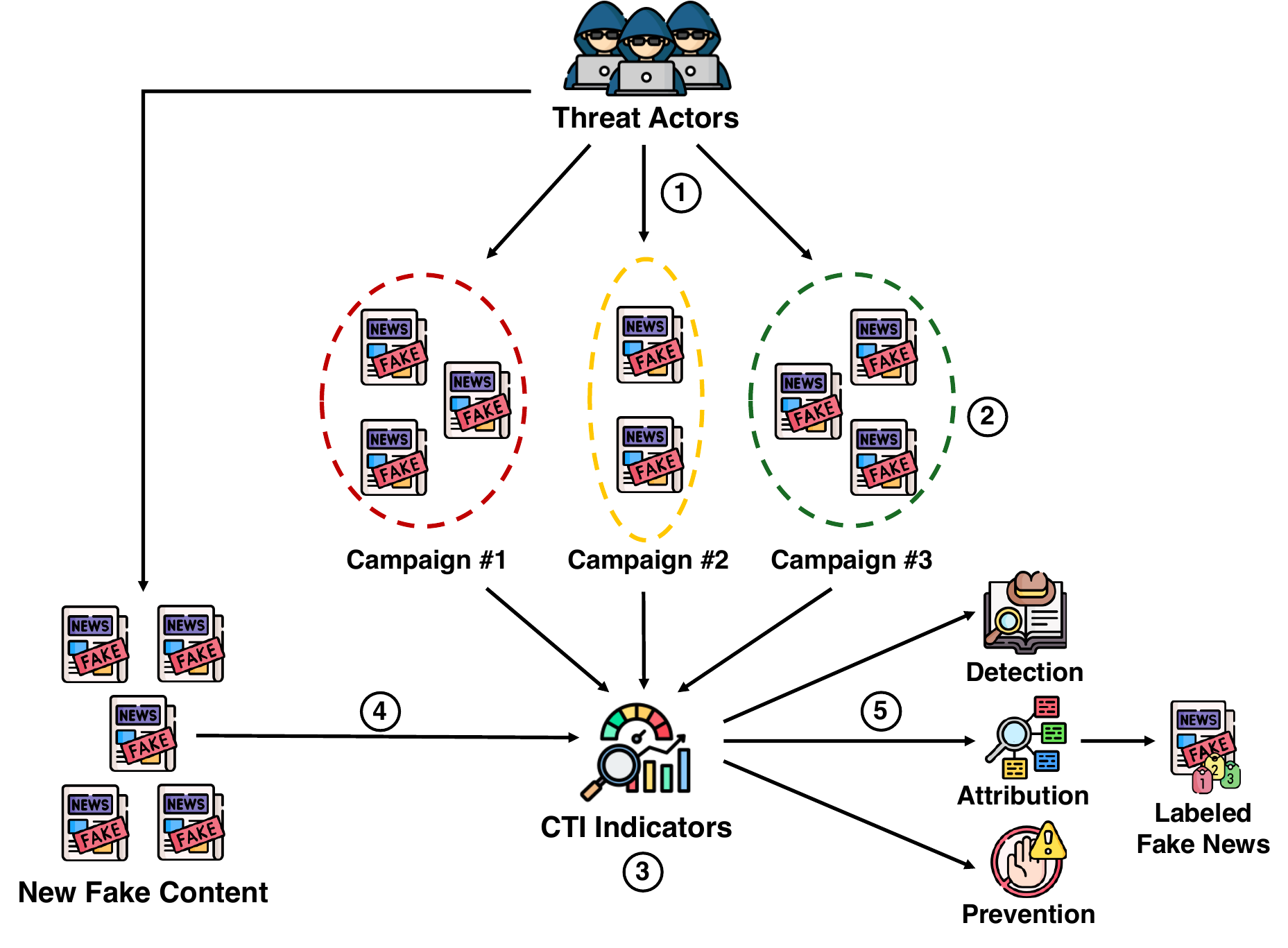}
  \caption{Overview of the proposal.}
  \label{fig:proposal}
\end{figure*}


Our work aims to define and extract concept-based Cyber Threat Intelligence (CTI) indicators to enhance the attribution and detection of fake news articles linked to disinformation campaigns. Figure \ref{fig:proposal} illustrates how our proposal is integrated into the disinformation landscape. Disinformation campaigns are orchestrated efforts by threat actors (\circled{1}) to spread false or misleading information to manipulate public opinion, influence political processes, or discredit individuals or institutions. These actors range from state-sponsored groups to independent propagandists and malicious organizations, leveraging various digital platforms to amplify their narratives. The fake news articles they produce form the foundation of these campaigns and are carefully crafted to blend with legitimate news, making detection and attribution increasingly challenging. These articles and social media posts are variants of the same fake content, spread using various web domains and social network accounts and crafted by rephrasing content.


Once published, these fake news articles become part of larger disinformation campaigns (\circled{2}), each characterized by distinct themes, narratives, and objectives. For example, a campaign may focus on public health disinformation, promoting anti-vaccine content, while another targets political elections, spreading conspiracy theories about electoral fraud. Establishing clear associations between individual fake news articles and their respective campaigns is crucial for attribution. However, this step remains difficult due to the lack of structured, high-level intelligence indicators that describe the narrative and intent behind the disinformation.

Traditional CTI approaches rely on low-level Indicators of Compromise (IoCs) such as IP addresses, domain names, and email addresses. While these indicators can help track technical infrastructure, they do not capture the semantic and conceptual features of disinformation. This gap necessitates the development of higher-level, content-driven CTI indicators (\circled{3}) that can characterize fake news articles based on their key claims, entities, and relationships rather than ephemeral technical artifacts.
To address this challenge, we introduce a new format of concept-based CTI indicators derived from the textual content of fake news articles. Unlike traditional IoCs, these indicators remain persistent across different campaigns and dissemination channels, making them more effective for long-term tracking and analysis.

Once novel CTI indicators have been defined, the next step is their automatic extraction from newly encountered fake news articles (\circled{4}). These structured intelligence indicators can help categorize, attribute, and track disinformation campaigns.

The extracted CTI indicators serve multiple purposes (\circled{5}), primarily in the following areas:
\begin{itemize}
    \item \textit{Detection}: concept-based CTI indicators enable the identification of new fake news articles that share similarities with known disinformation narratives. This allows early intervention before misleading information spreads widely.
    \item \textit{Attribution}: by linking extracted indicators to specific disinformation campaigns, we can trace which actors are responsible for spreading certain types of misinformation, improving situational awareness.
    \item \textit{Prevention}: identifying recurring patterns in disinformation helps in designing countermeasures to mitigate the impact of future campaigns, such as improving fact-checking mechanisms and enhancing automated content moderation.
\end{itemize}

\subsection{Concept-based CTI for fake news}
We propose a novel approach to structuring CTI indicators specifically tailored for fake news attribution. We focus on extracting meaningful, content-based indicators by analyzing the core claims, narratives, and references embedded in fake news articles. These indicators remain stable even as superficial details change, making them more effective for long-term attribution.
Such indicators should meet the following properties:
\begin{itemize}

    \item \textit{Non-volatile}: CTI indicators should reflect the contents of a disinformation campaign, without relying on technical artifacts that can be easily changed (domain names, social media accounts), and robust against variants of the same campaign.

    \item \textit{Interpretable}: CTI must be easily understood by both humans and automated systems, facilitating swift analysis and response.

    \item \textit{Suitable for supporting fake news detection}: CTI should be effective in guiding the detection of additional variants of fake news, enabling the identification of disinformation campaigns based on recognized patterns.

    \item \textit{Lightweight}: CTI should be quick to share and easy to store and process, ensuring efficient dissemination among stakeholders and minimal resource demand, in order to scale well for large volumes of data to analyze.

\end{itemize}

Consequently, we design indicators that take into account the underlying claims, narrative, and references of fake news articles, without relying on standard low-level information. Traditional CTI indicators, such as domain names, IP addresses, or social media account handles, are easily altered by threat actors, limiting their long-term usefulness for tracking disinformation campaigns.

Since fake news articles include explicit claims, implied narratives, and references to key entities, to facilitate their use for attribution tasks, we extract the following information:
\begin{itemize}
    \item \textit{Entities}: key actors, organizations, events, or concepts mentioned in the article.
    \item \textit{Relations}: the semantic connections between these entities. These relationships provide contextual meaning to the extracted entities.
    \item \textit{Objects}: the targets or outcomes associated with the entities and relations.
\end{itemize}


\begin{figure*}[htbp]
    \centering
    \includegraphics[width=\linewidth]{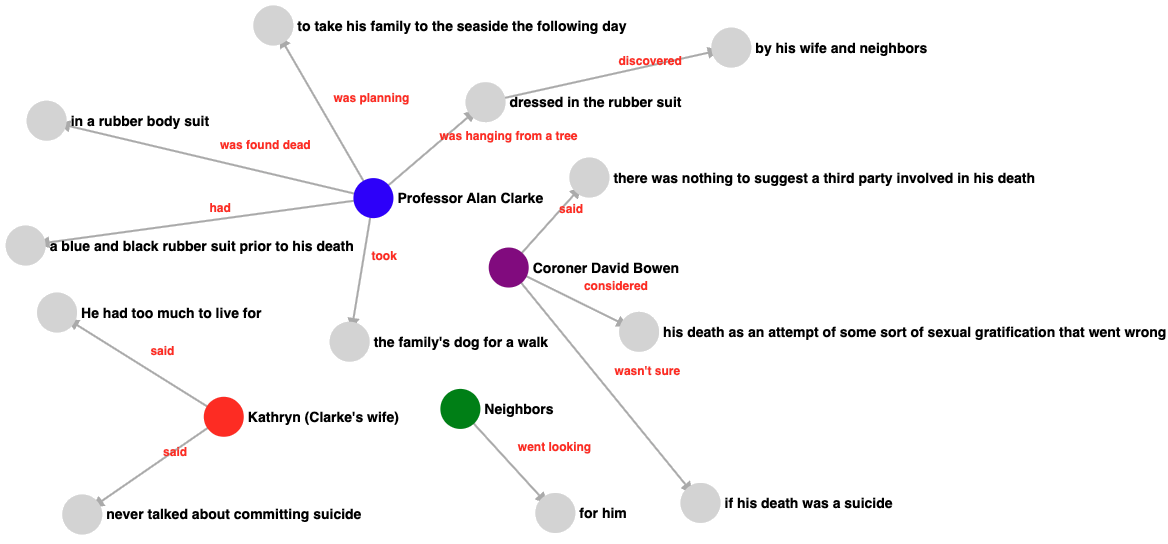}
    \caption{Example of graph extracted from a fake news article.} 
    \label{fig:Tuple-based graph}
\end{figure*}

This information is then grouped into \texttt{<subject, relation, object>} triples. In the following, we will refer to such triples as \textit{tuples}. 
Tuples are a widely adopted method for extracting semantic content from unstructured text \citep{stanovsky2018supervised, angeli2015leveraging, wang2021zero}, particularly in domains such as software engineering and log analysis \citep{ekelhart2021slogert, ladeinde2023extracting}.

For example, the sentence ``Country X funds Organization Y to spread misinformation" is transformed into the tuple \texttt{<Country X, funds, Organization Y to spread misinformation>} to retain the full semantic intent within a single tuple when possible. However, in cases where a sentence expresses multiple related actions or entities, it may be more appropriate to decompose it into multiple, more focused tuples to preserve informational fidelity and avoid under-specification. In this case, the original sentence can also be represented by two tuples: \texttt{<Country X, funds, Organization Y>} and \texttt{<Organization Y, spreads, misinformation>}, which explicitly separate the funding relationship from the resulting disinformation activity.

By capturing these core elements, we reduce the complexity of the text while preserving its critical meaning. The primary purposes of this process are i) to structure the unorganized, narrative-based content of fake news into a consistent format, ii) to enable the comparison of content across articles by focusing on entities and relationships, ignoring superficial variations in wording or phrasing, and iii) to identify recurring narratives, themes, and strategies within and across disinformation campaigns.

Although tuples may not preserve 100\% of the original textual nuance, their use still provides several advantages for attribution tasks.

One key benefit is their \textbf{non-volatile} nature. Unlike traditional indicators of compromise (IoCs), which threat actors can easily alter, tuples focus on the underlying content and intent of a disinformation campaign. The narrative or relationships between entities often remain consistent even as surface-level details change. For instance, while a campaign may migrate from one website to another, the core narrative captured by a tuple like \texttt{<Organization X, spreads, anti-vaccine propaganda>} is likely to persist.

Another significant advantage is \textbf{high interpretability}. Tuples offer a structured representation of information that is easily understandable for both human analysts and automated systems. A tuple such as \texttt{<Organization X, collaborates with, Country Y>} directly conveys the essence of the information, facilitating faster decision-making and more effective response planning.

Additionally, tuples enable \textbf{relationship identification} through mapping into a \textbf{relationship graph}, where nodes represent entities and edges define relationships. Such graphs allow for advanced analysis by highlighting recurring connections between entities, such as a particular actor repeatedly spreading similar narratives. They also reveal clusters of related campaigns or narratives, providing insights into potential coordination between different disinformation sources.

Figure \ref{fig:Tuple-based graph} illustrates an example of a graph extracted from a fake news article. Specifically, this example refers to an article discussing the mysterious death of Professor Alan Clarke \citep{AlanClarkeFakeNews}, a prominent UK cancer scientist and director of the European Cancer Stem Cell Research Institute. The discovery was made by his wife, Kathryn, and neighbors, who went looking for him after he failed to return from walking the family dog. The coroner, David Bowen, ruled out the involvement of a third party and was uncertain if Clarke's death was a suicide. Instead, Bowen speculated it might have been an attempt at sexual gratification that resulted in misadventure. 
The narrative ties into a disinformation campaign attributed to the threat actor \textit{Elizabeth Erin}, aiming to fuel distrust in the medical community by falsely linking the deaths of medical professionals to their work on cancer discoveries. The article additionally hints at broader conspiratorial undertones by referencing over 30 ``mysterious deaths" of prominent doctors, aligning with a known disinformation campaign titled \textit{Doctors Found Dead After Cancer Discovery}. It frames the situation with speculative and sensational details, such as pointing out Clarke's outfit and suggesting ulterior motives behind the coroner's public release of intimate information about the death.
The graph visualizes the relationships and facts presented in the article, connecting key subjects, actions, and objects such as Clarke’s attire, the coroner's verdict, and statements from witnesses.

\subsection{Tuples Extraction}
To automatically extract the tuples from a given fake news article, we leverage Large Language Models (LLMs). LLMs represent the most recent advancement in state-of-the-art relation extraction from text \citep{xu2024large, li2023revisiting, wan2023gpt}, making them the optimal choice for structuring information from unstructured sources like fake news articles. 
The effectiveness of LLM-based tuple extraction largely depends on the prompt structure and content. In our approach, we carefully design the prompt to provide clear and structured instructions, following best practices to ensure high-quality tuple extraction. The prompt includes a role description along with an explicit task description, according to prompt engineering best practices \citep{chen2023unleashing, white2023prompt}, guiding the model to focus on the extraction of \texttt{<subject, relation, object>} tuples while avoiding irrelevant details. Additionally, we define step-by-step extraction criteria to enhance consistency and reduce ambiguity in tuple formation. To improve interpretability, we include an example within the prompt, demonstrating the expected input-output structure. Finally, we append a termination keyword (\texttt{END LIST}) to ensure that the model correctly delimits its output, preventing unnecessary hallucination or extraneous text generation. This structured prompting methodology is designed to be model-agnostic, meaning it can be applied to any LLM capable of relation extraction. By following a clear role definition, explicit task description, structured step-by-step criteria, and a termination keyword, the prompt ensures consistency and accuracy across different models. Whether used with open-source or fine-tuned models, this approach remains adaptable, making it versatile for extracting structured information from unstructured text sources \cite{kim2023better, chen2025aecr}.


\begin{mainbox}{}
\textbf{Role:} You are a natural language processing expert specialized in analyzing textual data and extracting structured information. Your task is to identify subject-relation-object relationships from the input text, which represent key actions and relationships between entities. These tuples will be used for structured representation of fake news articles.

\textbf{Context:} Subject-relation-object relationships capture the fundamental structure of actions and relationships described in a sentence. In these relationships, the subject represents the entity performing an action, the verb describes the action or relationship, and the object represents the entity affected by the action. Identifying these relationships helps in organizing unstructured textual data into a structured format, enabling easier analysis and interpretation.

\textbf{Example:} Text: ``John gave a book to Mary.''\\
\phantom{\textbf{Example:} }Tuple: John - gave - a book to Mary

\textbf{Instructions:} Read the following text and identify all the tuples in the subject-verb-object form. The tuples should reflect the main actions and relationships between the entities mentioned in the text. Follow these steps:
\begin{enumerate}
    \item Identify the subject of the action.
    \item Identify the verb that describes the action or relationship.
    \item Identify the object or destination of the action.
\end{enumerate}
Return the tuples in this format: \textbf{Subject - Relation - Object}. At the end of the process, print \textbf{``END LIST"} to indicate the conclusion of the extraction.
\end{mainbox}


\subsection{Fake News Attribution}
\label{subsec:fake_news_attribution}
The final phase of the methodology uses concept-based CTI indicators to attribute newly detected fake news articles to previously identified campaigns. Thus, our approach maps an article to a known disinformation campaign and the threat actor behind it. The primary goal is to determine whether a piece of fake news belongs to an existing campaign and can be linked to specific threat actors. 
To this end, we employ both traditional techniques, such as lexical and semantic similarity metrics, and more advanced, LLM-based models to evaluate attribution performance. While traditional methods are widely considered vulnerable to paraphrasing and surface-level rewording, we include them in our analysis to serve as baselines and to investigate how they perform when applied to structured semantic representations (i.e., extracted \texttt{<subject, relation, object>} tuples) rather than unstructured raw text. Additionally, our evaluation is framed in the context of disinformation attribution, a more challenging and underexplored task compared to fake news detection, allowing us to assess whether structured content can enhance the robustness and generalizability of even simple methods. Specifically, we consider the following techniques:

\begin{itemize}
    \item \textit{Syntactic similarity}: detects direct or near-identical matches between articles, measuring the overlap in vocabulary and phrases between potential disinformation content and known fake news. To analyze the syntactic similarity between fake news articles and previously identified disinformation campaigns, we employed \textbf{Term Frequency-Inverse Document Frequency} (TF-IDF), widely used in NLP to assess the importance of words within a document relative to a larger corpus \cite{qaiser2018text, ramos2003using}. In this context, TF-IDF allows us to measure syntactic similarity between extracted \texttt{<subject, relation, object>} tuples and reference campaign data, enabling attribution based on textual overlap. To formalize this analysis for fake news attribution, we define the process of matching extracted tuples with those in the ground truth and assigning a campaign based on similarity. Let \( T = \{t_1, t_2, \dots, t_n\} \) be the set of extracted tuples from a fake news article under analysis, and \( C = \{C_1, C_2, \dots, C_m\} \) be the set of known disinformation campaigns, each characterized by a set of reference tuples \( T_{C_j} \).

    Each tuple \( t_i \) is transformed into a TF-IDF vector representation \( \mathbf{v}_i \). Similarly, each campaign \( C_j \) is represented by the TF-IDF matrix \( M_{C_j} \), where each row corresponds to the TF-IDF vector of a reference tuple associated with the campaign. The cosine similarity between an extracted tuple \( t_i \) and a campaign \( C_j \) is computed as:

    \begin{equation}
        \text{sim}(t_i, C_j) = \max_{t_k \in T_{C_j}} \frac{\mathbf{v}_i \cdot \mathbf{v}_k}{\|\mathbf{v}_i\| \|\mathbf{v}_k\|}
    \end{equation}

    where \( \mathbf{v}_i \) is the TF-IDF vector representation of the extracted tuple \( t_i \), \( \mathbf{v}_k \) is the TF-IDF vector representation of a reference tuple \( t_k \) from campaign \( C_j \), and \( \|\mathbf{v}_i\| \) and \( \|\mathbf{v}_k\| \) are the norms of the respective vectors. The maximum operator ensures that the best-matching tuple in the campaign is considered.

    A tuple \( t_i \) is assigned to a campaign \( C_j \) if:

    \begin{equation}
        \text{sim}(t_i, C_j) \geq \tau
    \end{equation}

    where \( \tau \) is a predefined similarity threshold. If no campaign satisfies this condition, the tuple remains unassigned.

    For each fake news article, a \textbf{voting mechanism} determines the most probable campaign based on the number of tuples assigned to each campaign. Let \( V_{C_j} \) be the vote count for campaign \( C_j \), computed as:

    \begin{equation}
        V_{C_j} = \sum_{t_i \in T} 1(\text{sim}(t_i, C_j) \geq \tau)
    \end{equation}

    where \( 1(\cdot) \) is the indicator function:
    \begin{equation}
        1(\text{sim}(t_i, C_j) \geq \tau) =
        \begin{cases}
            1, & \text{if } \text{sim}(t_i, C_j) \geq \tau \\
            0, & \text{otherwise}
        \end{cases}
    \end{equation}

    The article is ultimately assigned to the campaign with the highest vote count:
    \begin{equation}
        C^* = \arg\max_{C_j \in C} V_{C_j}
    \end{equation}
    where \( C^* \) represents the most probable disinformation campaign associated with the article.
    
    \item \textit{Semantic similarity}: leverages \textit{word embeddings} to capture the meaning of the content beyond the exact wording. This approach allows for identifying nuanced connections between potentially misleading news and known disinformation content. This analysis follows a similar approach as in the lexical similarity analysis but with the primary distinction of using semantic embeddings instead of lexical frequency-based representations.
    Let \( T = \{t_1, t_2, \dots, t_n\} \) be the set of extracted tuples from a fake news article, and let \( C = \{C_1, C_2, \dots, C_m\} \) be the set of known disinformation campaigns, each characterized by a set of reference tuples \( T_{C_j} \).
    Each tuple \( t_i \) is transformed into a semantic embedding \( \mathbf{e}_i \) in a high-dimensional vector space. Similarly, each campaign \( C_j \) is represented by an embedding matrix \( M_{C_j} \), where each row corresponds to the embedding of a reference tuple associated with the campaign.

    The cosine similarity between an extracted tuple \( t_i \) and a campaign \( C_j \) is computed as:
    
    \begin{equation}
        \text{sim}(t_i, C_j) = \max_{t_k \in T_{C_j}} \frac{\mathbf{e}_i \cdot \mathbf{e}_k}{\|\mathbf{e}_i\| \|\mathbf{e}_k\|}
    \end{equation}
    where \( \mathbf{e}_i \) is the embedding representation of the extracted tuple \( t_i \), \( \mathbf{e}_k \) is the embedding representation of a reference tuple \( t_k \) from campaign \( C_j \), and \( \|\mathbf{e}_i\| \) and \( \|\mathbf{e}_k\| \) are the norms of the respective embedding vectors.

    A tuple \( t_i \) is assigned to a campaign \( C_j \) if:
    \begin{equation}
        \text{sim}(t_i, C_j) \geq \tau
    \end{equation}
    where \( \tau \) is a predefined similarity threshold. If no campaign satisfies this condition, the tuple remains unassigned.
    Again, for each article, a voting mechanism determines the most probable campaign based on the number of tuples assigned to each campaign. Let \( V_{C_j} \) be the vote count for campaign \( C_j \), computed as:
    \begin{equation}
        V_{C_j} = \sum_{t_i \in T} 1(\text{sim}(t_i, C_j) \geq \tau)
    \end{equation}
    where \( 1(\cdot) \) is the indicator function:
    \begin{equation}
        1(\text{sim}(t_i, C_j) \geq \tau) =
        \begin{cases}
            1, & \text{if } \text{sim}(t_i, C_j) \geq \tau \\
            0, & \text{otherwise}
        \end{cases}
    \end{equation}

    The article is ultimately assigned to the campaign with the highest vote count:
    \begin{equation}
        C^* = \arg\max_{C_j \in C} V_{C_j}
    \end{equation}
    where \( C^* \) represents the most probable disinformation campaign associated with the article.
    
    \item \textit{Large Language Models}: this approach uses machine learning approaches based on LLMs that are fine-tuned to recognize campaign-specific language patterns and relationships between entities. These models classify fake news articles by disinformation campaign and threat actor. Specifically, we leverage DistilBERT, a lighter, faster variant of BERT, designed to retain most of its capabilities while reducing computational overhead through knowledge distillation \citep{Sanh2019DistilBERTAD}. DistilBERT retains the core transformer architecture of BERT, relying on \textit{self-attention mechanisms} and \textit{feed-forward layers} to process and understand text. It follows the standard encoder-only transformer structure, where input text is tokenized and transformed into vector representations before passing through multiple transformer layers.

    The self-attention mechanism \cite{vaswani2017attention}, a key component of transformers, computes contextual relationships between words in a sequence. Given an input sentence tokenized into words \(\{ w_1, w_2, \dots, w_n \}\), each word is transformed into a high-dimensional embedding vector. The transformer then applies multi-head self-attention to compute contextualized representations. The self-attention mechanism computes a weighted sum of all words in the input sequence to determine how much attention each word should pay to others. This is achieved using three matrices: Query (\(Q\)), Key (\(K\)), and Value (\(V\)).

    For each word token \(x_i\), these matrices are computed as:
    \begin{equation}
        Q = XW_Q, \quad K = XW_K, \quad V = XW_V
    \end{equation}
    where \( X \in \mathbb{R}^{n \times d} \) is the input matrix (word embeddings), \( W_Q, W_K, W_V \in \mathbb{R}^{d \times d_k} \) are learnable weight matrices, \( d \) is the embedding dimension, and \( d_k \) is the dimensionality of queries and keys. The attention score between two tokens is computed as:
    \begin{equation}
        \text{Attention}(Q, K, V) = \text{softmax} \left( \frac{Q K^T}{\sqrt{d_k}} \right) V
    \end{equation}
    where \( Q K^T \) computes the similarity between queries and keys, \( \sqrt{d_k} \) scales the values to stabilize training, and the softmax function normalizes the scores so they sum to 1. This allows the model to focus on the most relevant parts of the input sentence.
    DistilBERT maintains multi-head attention, where multiple independent self-attention computations are performed in parallel, allowing the model to capture different types of relationships between words.

    The goal is to let the model learn how to associate extracted tuples with their corresponding disinformation campaigns based on their semantic content. Given a set of fake news articles \( A = \{a_1, a_2, \dots, a_n\} \) and a predefined set of disinformation campaigns \( C = \{C_1, C_2, \dots, C_m\} \), the model's task is to classify each article \( a_i \) by analyzing the extracted tuples.

    Each article \( a_i \) consists of a set of tuples \( T = \{t_1, t_2, \dots, t_n\} \) where each tuple \( t_j \) represents structured information in the \texttt{<subject, relation, object>} form. These tuples serve as the input to the model.

    To process the input, each tuple \( t_j \) is tokenized and transformed into a numerical representation \( X_j \):
    \[
        X_j = f_{\text{tokenizer}}(t_j)
    \]
    where \( f_{\text{tokenizer}}(\cdot) \) maps the tuple into a tokenized sequence compatible with DistilBERT. The model is trained to predict the most probable campaign label \( C_j \) for each tuple \( t_j \).

    For each input tuple \( t_j \), DistilBERT produces a probability distribution over the possible disinformation campaigns:
    \[
        P(C | t_j) = \text{softmax}(W \mathbf{h}_{\text{CLS}} + b)
    \]
    where \( W \) is a learned weight matrix, \( b \) is a bias term, \( \mathbf{h}_{\text{CLS}} \) is the output representation of the DistilBERT classification token, and \( P(C | t_j) \) represents the predicted probability distribution over campaigns.
    The campaign assigned to tuple \( t_j \) is:
    \[
        C^*(t_j) = \arg\max_{C_j} P(C_j | t_j)
    \]
    where \( C^*(t_j) \) is the campaign with the highest predicted probability.

    Since each fake news article \( a_i \) consists of multiple tuples, we apply a majority voting mechanism to determine its final campaign classification. Let \( V_{C_j} \) be the vote count for campaign \( C_j \):
    \[
        V_{C_j} = \sum_{t_j \in T_i} 1(C^*(t_j) = C_j)
    \]
    where \( 1(\cdot) \) is the indicator function:
    \[
        1(C^*(t_j) = C_j) =
        \begin{cases}
            1, & \text{if } C^*(t_j) = C_j \\
            0, & \text{otherwise}
        \end{cases}
    \]
    The final campaign assignment for the article \( a_i \) is given by:
    \[
        C^*(a_i) = \arg\max_{C_j} V_{C_j}
    \]
    ensuring that the campaign with the most assigned tuples determines the article’s classification.
\end{itemize}

These three solutions differ in terms of performance, cost, and accuracy. Syntactic similarity is computationally efficient and straightforward but may fail to capture deeper semantic relationships between articles. Semantic similarity achieves a better understanding of content meaning by utilizing word embeddings, albeit at a higher computational cost. Fine-tuned LLMs offer the highest accuracy and flexibility in identifying complex patterns and relationships. However, their increased computational cost and the requirement for high-quality training data make them the most resource-intensive option. 


\section{The FakeCTI Dataset}
\label{sec:dataset}

To support our investigation of concept-based CTI indicators, we created a new collection of fake news content in the \textit{FakeCTI} dataset. This structured dataset facilitates the study of disinformation campaigns, grouping fake news that belong to the same campaign and associating the campaigns with threat actors.

We gathered data from diverse and scattered sources linked by Wikipedia in the following archives:
\begin{itemize}
    \item \textit{Political disinformation website campaigns in the United States} \citep{WikipediaFakeNews1};
    \item \textit{Fake news websites} \citep{WikipediaFakeNews2};
    \item \textit{Fake news troll farms} \citep{WikipediaFakeNews3};
    \item \textit{Corporate disinformation website campaigns} \citep{WikipediaFakeNews4}.
\end{itemize}

These archives collect information about several fake news campaigns, in the form of unstructured Cyber Threat Intelligence (CTI) reports \citep{CTIExample1, CTIExample2, CTIExample3}. The campaigns have been categorized into various domains such as political disinformation campaigns, troll farms, fake news websites, and corporate disinformation efforts. These reports document the origins, methods, and motivations of fake news, providing insights into how it is created and disseminated.

Political disinformation campaigns often stem from partisan entities or super PACs, with websites like American Action News designed to sway public opinion during elections by spreading conspiracy theories or ideologically charged narratives. Fake news websites, such as Conservative Beaver and The Red Panther, use deceptive tactics like spoofed domains and exaggerated clickbait to generate traffic and ad revenue, typically driven by independent networks or profit-focused individuals. Troll farms, as seen in operations from Ghana or Myanmar, use coordinated networks of sites to amplify divisive or false stories, frequently backed by anonymous contributors or political actors aiming to manipulate public discourse. Corporate disinformation efforts, exemplified by platforms like Chevron-funded The Richmond Standard, masquerade as community news outlets to promote favorable narratives and suppress criticism, often managed by PR firms.

\begin{figure}[htbp]
    \centering
    \includegraphics[width=\linewidth]{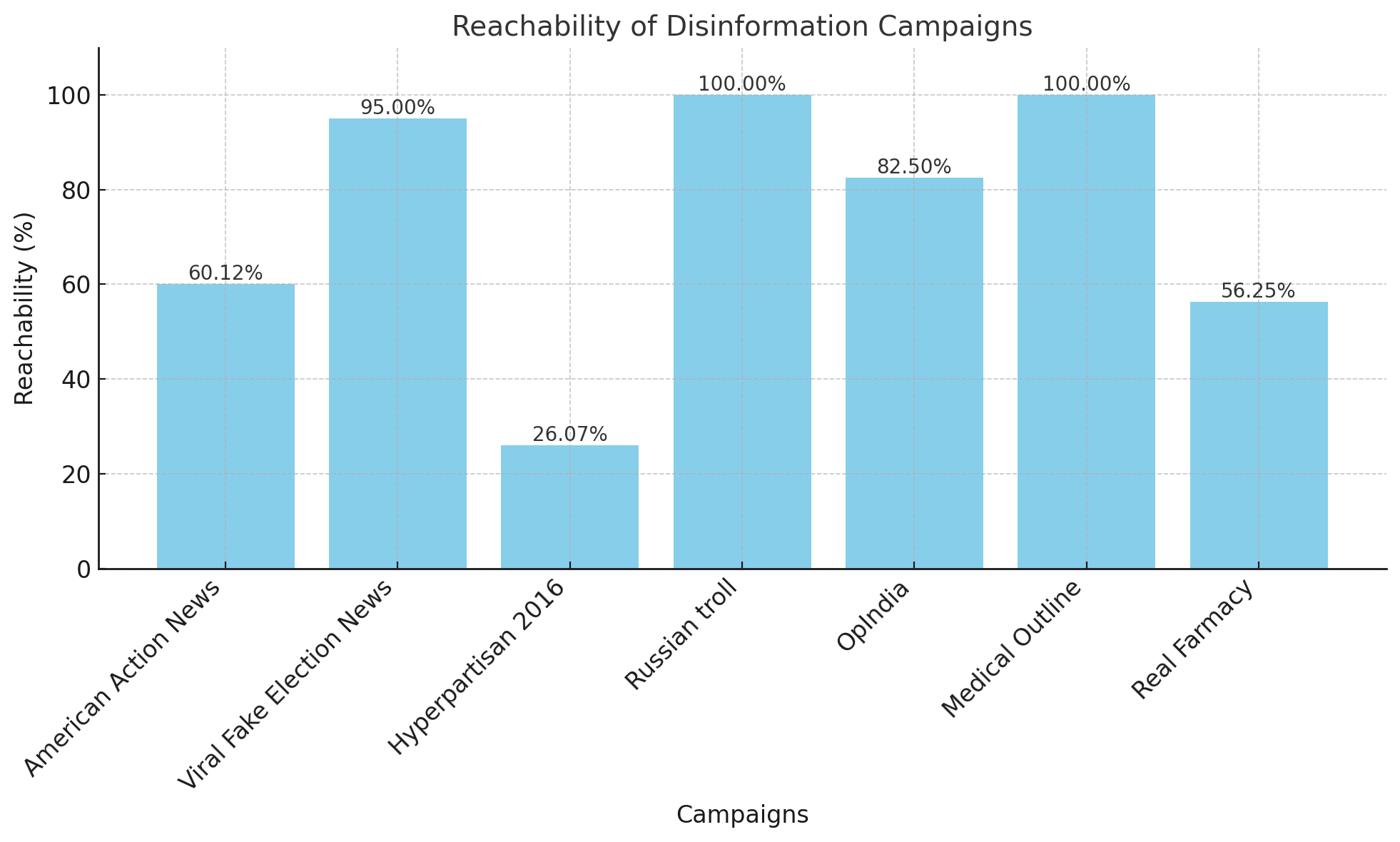}
    \caption{Reachability of disinformation campaigns.}
    \label{fig:Campaign reachability}
\end{figure}

To create the FakeCTI dataset, we extracted data from the websites and platforms used to disseminate fake news and referenced in these archives and systematized the data to make them suitable for research studies. 
After collecting the raw data from these sources, we removed irrelevant or low-quality entries, such as duplicate articles. Then, using the metadata extracted from the CTI reports, we manually labeled the disinformation articles based on their association with specific disinformation campaigns and threat actors. 
Finally, we linked fake news to the channels used to spread it, such as social media platforms, blogs, and news websites.

A challenge in building the dataset is the reachability of the disinformation campaigns. Not all fake news articles included in the original campaigns were accessible during the data collection process. Many articles were no longer available, even through the use of web archives that capture historical snapshots of web pages. This limitation arises because mainstream web portals and information agencies often remove false content once it is reported, as part of their efforts to protect readers from disinformation. As a result, a limitation of the dataset is that it reflects only the accessible portion of the content of the campaigns.

Figure \ref{fig:Campaign reachability} shows the reachability of contents for various disinformation campaigns in the dataset, with respect to the total number of articles reported for the campaigns. The reachability metric quantifies the proportion of articles from each campaign that could still be accessed and retrieved during the data collection process. Campaigns such as \textit{Russian Troll} and \textit{Medical Outline} exhibit full reachability (100\%), indicating that all known articles from these campaigns were accessible at the time of writing. Other campaigns, such as \textit{Hyperpartisan 2016} and \textit{Real Farmacy}, have significantly lower reachability at 26.07\% and 56.25\%, respectively, reflecting the challenges in retrieving articles due to factors like website takedowns or insufficient archival coverage. This variability highlights a limitation in the dataset's completeness, as campaigns with lower reachability may be underrepresented.


Despite the limitations, we obtained a significant amount of data on fake news campaigns. Table \ref{tab:dataset_stats} shows some statistics about the FakeCTI dataset. The dataset contains a total of 12,155 fake news samples, encompassing 43 distinct disinformation campaigns orchestrated by 73 identified threat actors. FakeCTI aggregates content from 149 different sources, offering a diverse and comprehensive foundation for studying the dissemination and attribution of fake news across multiple campaigns and actors.

\begin{table}[htbp]
\caption{Summary of FakeCTI dataset statistics.}
\centering
\begin{tabular}{lc}
\toprule
\textbf{Statistic} & \textbf{Value} \\
\midrule
Total number of samples      & 12,155 \\
Number of different threat actors & 73 \\
Number of different fake news campaigns & 43 \\
Number of different sources  & 149 \\
\bottomrule
\end{tabular}
\label{tab:dataset_stats}
\end{table}

\begin{table*}[htbp]
\caption{Example of FakeCTI dataset entries.}
\begin{center}
\resizebox{\linewidth}{!}{%
\begin{tabular}{p{0.35\textwidth}>
{\centering}p{0.15\textwidth}>
{\centering}p{0.2\textwidth}>
{\centering}p{0.2\textwidth}>
{\centering\arraybackslash}p{0.23\textwidth}}
    \toprule
    \textbf{Title} & \textbf{Source} & \textbf{Campaign} & \textbf{Threat Actor} & \textbf{Medium}\\
    \midrule
    Vaccinated people are walking biological time bombs and a THREAT to society & Foreign Affairs Intelligence Council & Covid trigger Neurological Degenerative Diseases & Alex Jones & Web \\
    \midrule
    LEAKED: Lady Gaga Halftime Performance to Feature Muslim Tribute & Houston Leader & A Cure for Wellness Promotion & Regency Enterprise & Web \\
    \midrule
    Pope Francis Shocks World, Endorses Donald Trump for President, Releases Statement & Ending the Fed News & Viral Fake Election News & Unknown & Web \\
    \bottomrule
\end{tabular}}
\label{tab:Dataset Examples}
\end{center}
\end{table*}

The resulting entries in the FakeCTI dataset include the following information: the \textit{title} and \textit{text} of the fake news article, a \textit{link} to the article itself, the \textit{disinformation campaign} that spread the fake news, the \textit{threat actor} behind the campaign, and the \textit{dissemination medium} used for spreading fake news. Table \ref{tab:Dataset Examples} illustrates some sample entries from the dataset. The first entry originates from the Foreign Affairs Intelligence Council as part of the campaign \textit{Covid triggers Neurological Degenerative Diseases} attributed to Alex Jones, a known figure in spreading health-related disinformation. The second article was published by Houston Leader and is associated with the campaign \textit{A Cure for Wellness Promotion} with its threat actor identified as Regency Enterprise, likely linked to a promotional stunt. Lastly, the third example comes from Ending the Fed News and is part of the \textit{Viral Fake Election News} campaign, which refers to the 2016 presidential election in the United States. It is possible to notice how this entry lacks an identified threat actor. This absence highlights that such attribution is not always possible due to the anonymous and decentralized nature of many disinformation efforts.

This dataset bridges a significant gap in existing resources, providing a richer context for fake news attribution tasks. It allows analysts to link disinformation narratives to their broader campaigns and threat actors, making the dataset an invaluable tool for attribution and detection research.

\begin{figure*}[ht]
\centering
\includegraphics[width=0.9\linewidth]{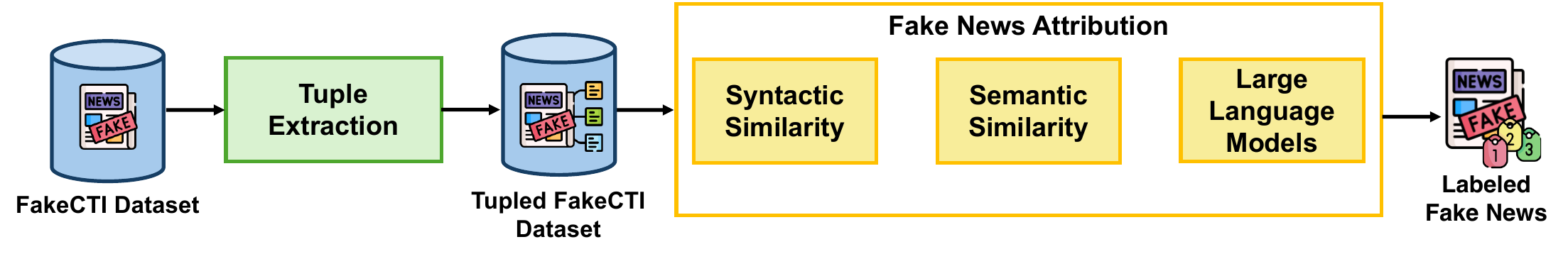}
  \caption{Overview of the experimental analysis.}
  \label{fig:evaluation}
\end{figure*}

\section{Research Questions}
\label{sec:questions}
We evaluated the effectiveness of the proposed approach by considering the following research questions (RQ).

\vspace{2mm}
\noindent
$\rhd$ \textbf{RQ1}: \textit{How accurately can LLMs extract tuples from fake news articles, compared to manually annotated ground truth?}\\
RQ1 investigates the quality of the tuples generated by LLMs in the context of disinformation analysis. The goal is to evaluate the quality of the extracted tuples against a manually curated dataset, ensuring that the LLMs can capture the critical relationships and information in fake news articles. This assessment is crucial for determining whether LLMs can reliably standardize and structure unstructured textual data for automated analysis.

\vspace{2mm}
\noindent
$\rhd$ \textbf{RQ2}: \textit{How effective are the extracted CTI indicators for attributing fake news articles to specific disinformation campaigns and threat actors?}\\
RQ2 examines whether the CTI indicators can accurately and comprehensively be used to link fake news articles to the appropriate disinformation campaigns or threat actors. By testing attribution techniques (e.g., lexical similarity, semantic similarity, and fine-tuned LLMs), we aim to validate whether the structured tuples sufficiently represent the key characteristics of the campaigns to support meaningful and precise attribution efforts. Given the broad focus of RQ2, we further refine this research question into sub-questions RQ2.1 and RQ2.2.

\vspace{2mm}
\noindent
$\rhd$ \textbf{RQ2.1}: \textit{How effectively can traditional NLP techniques associate extracted tuples with known disinformation campaigns?}\\
RQ2.1 aims to evaluate the effectiveness of standard NLP methods in mapping extracted \texttt{<subject, relation, object>} tuples to their respective disinformation campaigns. The focus is on assessing how well lexical and semantic similarity measures can establish meaningful connections between tuples and campaign narratives. 

\vspace{2mm}
\noindent
$\rhd$ \textbf{RQ2.2}: \textit{Can fine-tuned LLMs improve the accuracy and reliability of fake news attribution compared to traditional NLP-based approaches?}\\
RQ2.2 investigates whether fine-tuning transformer-based LLMs enhances the attribution process beyond traditional methods. The study examines whether these models, trained specifically on disinformation-related data, can capture more nuanced patterns in tuple-based representations and improve classification performance, ultimately leading to more robust and scalable fake news attribution.

\section{Experimental Analysis}
\label{sec:experiment}
To evaluate the effectiveness of CTI indicators, we conducted an experimental analysis of fake news attribution techniques on the \textit{FakeCTI} dataset. These experiments focused on identifying the disinformation campaign behind a given fake news article. Figure \ref{fig:evaluation} illustrates the workflow for the experimental evaluation. First, we start with \textit{tuple extraction} from the FakeCTI dataset. Then, leveraging the tupled version of the dataset, we test several techniques for \textit{fake news attribution} to assess the concept-based CTI indicators. 

\subsection{Tuple Extraction}
To extract structured information from fake news articles, we leveraged two state-of-the-art LLMs: Meta’s LLaMA and DeepSeek. We chose these models for their high performance in natural language understanding and their architectural suitability for information extraction tasks. LLaMA, specifically the distilled \texttt{Llama-3-8B-Instruct-v0.10} variant \citep{LlamaHuggingFace}, offers 8 billion parameters and excels at deep semantic parsing and context-aware generation, making it well-suited for capturing relational structures in complex narratives. Despite the efficiency improvements from distillation, executing LLaMA on available hardware resources, particularly GPU memory, remained challenging. To address this, we implemented quantization, a technique that reduces numerical precision in model parameters and calculations. While traditional LLMs use 32-bit floating point representations, quantization allows for reduced-precision formats (e.g., 8-bit integers), significantly lowering memory and computation requirements with minimal loss in inference accuracy, thanks to modern quantization algorithms \citep{lang2024comprehensive, wu2020integer}. In parallel, we also employed \texttt{DeepSeek-Coder-6.7B} \citep{DeepSeekHuggingFace}, a multilingual, instruction-tuned model known for its strong generalization and fine-grained reasoning capabilities, particularly within open-source ecosystems. We selected these two models to ensure a robust comparison between LLMs with distinct pre-training paradigms and optimization strategies, thereby evaluating the consistency and reliability of tuple extraction across different LLM architectures.

Specifically, we tested three quantized versions of LLaMA: Q4\_K\_M (4-bit quantization), Q5\_K\_M (5-bit quantization), and Q8\_0 (8-bit quantization). These configurations, along with DeepSeek, were evaluated based on execution time and precision, ensuring an optimal balance between efficiency and extraction accuracy.



We then proceeded with the extraction of \texttt{<subject, relation, object>} tuples from fake news articles. This process posed several challenges, including how to optimize the extraction process for better accuracy and how to evaluate the extracted results. For the LLaMA model specifically, a key factor influencing extraction quality was the \textit{temperature} parameter, which controls randomness in the model's responses. Low temperatures (close to 0) make responses more deterministic, ensuring consistent extractions but potentially limiting the model's flexibility in capturing diverse expressions, while higher temperatures (close to 1) increase variability, promoting diversity but at the risk of inconsistency or hallucination.
To investigate the impact of this parameter, we conducted a preliminary study and tested three temperature values: $0$ (fully deterministic), $0.3$ (moderate variability), and $0.6$ (higher variability). We performed this evaluation on a reduced subset of articles to assess how temperature settings influence the trade-off between extraction precision and generalization. Table \ref{tab:preliminary_extraction_results} illustrates the results of this analysis. It is possible to notice how the temperature of 0.3 consistently yields the highest precision, recall, and F1-Score, indicating an optimal balance between deterministic behavior and generation diversity. For instance, the Q4\_K\_M model at temperature 0.3 achieves the best overall performance with an F1-Score of 69\%, outperforming both lower (0.0) and higher (0.6) temperature settings. In contrast, while producing more deterministic outputs, the temperature 0.0 results in slightly lower recall, and temperature 0.6 tends to degrade performance across all metrics, likely due to increased generation variability and hallucinations. These findings justify the choice of 0.3 as the default temperature for the main evaluation, as it offers the most robust extraction quality while maintaining reasonable computational efficiency.


\begin{table*}[htbp]
\centering
\caption{Performance comparison across quantized LLaMA models and temperature settings for tuple extraction. Best performance is \textcolor{red}{\textbf{red}}.}
\begin{tabular}{cccccc}
    \toprule
    \textbf{Model} & \textbf{Temperature} & \textbf{Precision (\%)} & \textbf{Recall (\%)} & \textbf{F1-Score (\%)} & \textbf{Avg. Extraction Time (s)}\\
    \midrule
    \multirow{3}{*}{Q4\_K\_M} & 0.0 & 63 \DrawPercentageBar{0.63} & 65 \DrawPercentageBar{0.65} & 64 \DrawPercentageBar{0.64} & 25.07\\
    & \textcolor{red}{\textbf{0.3}} & \textcolor{red}{\textbf{67}} \DrawPercentageBarRed{0.67} & \textcolor{red}{\textbf{70}} \DrawPercentageBarRed{0.7} & \textcolor{red}{\textbf{69}} \DrawPercentageBarRed{0.69} & 29.71\\
    & 0.6 & 60 \DrawPercentageBar{0.6} & 63 \DrawPercentageBar{0.63} & 61 \DrawPercentageBar{0.61} & \textcolor{red}{\textbf{24.61}}\\
    \midrule
    \multirow{3}{*}{Q5\_K\_M} & 0.0 & 58 \DrawPercentageBar{0.58} & 61 \DrawPercentageBar{0.61} & 59 \DrawPercentageBar{0.59} & 26.14\\
    & \textcolor{red}{\textbf{0.3}} & \textcolor{red}{\textbf{61}} \DrawPercentageBarRed{0.61} & \textcolor{red}{\textbf{64}} \DrawPercentageBarRed{0.64} & \textcolor{red}{\textbf{62}} \DrawPercentageBarRed{0.62} & 31.53\\
    & 0.6 & 56 \DrawPercentageBar{0.56} & 60 \DrawPercentageBar{0.6} & 57 \DrawPercentageBar{0.57} & \textcolor{red}{\textbf{31.45}}\\
    \midrule
    \multirow{3}{*}{Q8\_0} & 0.0 & 54 \DrawPercentageBar{0.54} & 58 \DrawPercentageBar{0.58} & 55 \DrawPercentageBar{0.55} & 29.81\\
    & \textcolor{red}{\textbf{0.3}} & \textcolor{red}{\textbf{57}} \DrawPercentageBarRed{0.57} & \textcolor{red}{\textbf{60}} \DrawPercentageBarRed{0.6} & \textcolor{red}{\textbf{58}} \DrawPercentageBarRed{0.58} & \textcolor{red}{\textbf{23.60}}\\
    & 0.6 & 50 \DrawPercentageBar{0.5} & 54 \DrawPercentageBar{0.54} & 51 \DrawPercentageBar{0.51} & 44.79\\
    \bottomrule
\end{tabular}
\label{tab:preliminary_extraction_results}
\end{table*}



To evaluate the ability of the LLMs to extract meaningful and structured concepts from fake news articles, we conducted a systematic evaluation comparing model-generated tuples with human-annotated ground truth. The objective was to determine how effectively LLMs can represent disinformation content through structured \texttt{<subject, relation, object>} tuples, and how well these representations preserve the key semantic elements of the original articles.

We built the ground truth through the following steps: 
\begin{enumerate}
    \item \textit{Random selection of articles}: we randomly sampled a subset of 10 articles per campaign from the dataset to ensure diversity in the topic and linguistic style.
    \item \textit{Manual concept identification}: each selected article was read to manually identify a set of high-level concepts that best encapsulate the main disinformation themes. These concepts represent the essential assertions made in the article and serve as semantic anchors for subsequent evaluation. This step was carried out in collaboration with two MSc students with academic backgrounds in cybersecurity and Cyber Threat Intelligence, ensuring domain expertise in interpreting and annotating deceptive narratives. The authors then independently reviewed and cross-validated the extracted concepts, resolving ambiguities and harmonizing terminology through collaborative discussion to establish a unified and coherent annotation standard.
\end{enumerate}

We illustrate the concept extraction process using an article titled ``Arms for Ukraine ended up on black market in Finland", attributed to a pro-Kremlin disinformation campaign in the context of the Russian invasion of Ukraine. Table \ref{tab:concept_extraction_mapping} presents the mapping between article sentences and the manually extracted concepts.

To assess the quality of the tuples extracted by the two LLMs concerning the manually identified concepts, we defined two complementary evaluation metrics: \textit{Accuracy} and \textit{Coverage}. First, Accuracy measures whether each extracted tuple correctly and faithfully captures a factual or implied assertion present in the original article. We define this metric as:
\[
\text{\textit{Accuracy}} = \frac{\text{N}_{\text{CT}}}{\text{N}_{\text{ET}}}
\]
\noindent
where $\text{N}_{\text{CT}}$ is the number of correct tuples extracted from a specific article, while $\text{N}_{\text{ET}}$ is the total number of extracted tuples.
Second, Coverage determines whether each of the manually extracted concepts is matched by at least one semantically coherent tuple. This metric captures how comprehensively the model represents the full range of key narrative elements within the article and is defined as:
\[
\text{\textit{Coverage}} = \frac{\text{N}_{\text{CC}}}{\text{N}_{\text{TC}}}
\]
\noindent
where $\text{N}_{\text{CC}}$ is the number of concepts covered by the extracted tuples, given a specific article, while $\text{N}_{\text{TC}}$ is the total number of manually extracted concepts.\\
These two metrics are designed to jointly provide an exhaustive evaluation of tuple quality, as they capture both the fidelity of individual extractions to the source content and the completeness of the overall narrative representation. 

\begin{table*}[htbp]
\caption{Mapping of article sentences to manually extracted narrative concepts.}
\centering
\resizebox{0.85\textwidth}{!}{%
\begin{tabular}{p{0.48\textwidth} p{0.48\textwidth}}
\toprule
\textbf{Article Sentence} & \textbf{Extracted Concept} \\
\midrule
The weapons, which were intended for Ukraine, have reached Finland and ended up in the hands of organized crime. & Weapons destined for Ukraine reach Finland, end up with organized crime \\
\midrule
This was reported by Finnish media with reference to the Commissioner of the Central Criminal Police, Krister Ahlgren. & Finnish media cite Commissioner Krister Ahlgren with evidence \\
\midrule
“We have evidence that the weapons supplied to Ukraine are already in Finland. Weapons have also been found in Sweden, Denmark and the Netherlands,” the commissioner said. & Weapons also found in Sweden, Denmark, the Netherlands \\
\midrule
According to Ahlgren, the findings involve assault rifles in particular. & Finds are mostly for assault rifles \\
\midrule
More detailed information was not available for reasons of secrecy and ongoing investigation. & Confidential information omitted for secrecy and investigation \\
\midrule
Earlier, even Europol warned that uncontrolled supplies of weapons to Ukraine lead to the black market. & Europol had warned: uncontrolled supplies $\rightarrow$ black market \\
\midrule
And advertisements have begun to appear on the darknet for the sale of not just small arms, but even Javelin anti-tank missiles. & Darknet ads for Javelin small arms and missiles \\
\bottomrule
\end{tabular}
}%
\label{tab:concept_extraction_mapping}
\end{table*}

\begin{table}[htbp]
\centering
\caption{Performance comparison between LLMs for tuple extraction. Best performance is \textcolor{red}{\textbf{red}}.}
\resizebox{\columnwidth}{!}{%
\begin{tabular}{cccccc}
    \toprule
    \textbf{Model} & \textbf{Accuracy (\%)} & \textbf{Coverage (\%)} & \textbf{Avg. Extraction Time (s)}\\
    \midrule
    LLaMA & \textcolor{red}{\textbf{81.6}} \DrawPercentageBarRed{0.816} & 60.5 \DrawPercentageBar{0.605} & 25.1\\
    \midrule
    DeepSeek & 78.9 \DrawPercentageBar{0.789} & \textcolor{red}{\textbf{83.4}} \DrawPercentageBarRed{0.834} & \textcolor{red}{\textbf{8.5}} \\
    \bottomrule
\end{tabular}
}%
\label{tab:tuple_extraction_results}
\end{table}

Table \ref{tab:tuple_extraction_results} illustrates the results of the analysis. For LLaMA, we conducted the evaluation using the Q5\_K\_M quantized variant with a temperature of 0.3, as preliminary experiments in Table \ref{tab:preliminary_extraction_results} showed that quantization had no significant effect on tuple extraction quality across the articles and that the 0.3 was the best-suited value for the temperature parameter. 
LLaMA achieves the highest accuracy at 81.6\%, indicating its strong ability to generate precise and semantically faithful tuples that closely align with the original article content. This reflects LLaMA’s robust syntactic and semantic understanding, particularly for shorter or structurally clear sentences, where its deterministic behavior at low temperatures appears to minimize hallucinations and structural errors. However, LLaMA underperforms in terms of coverage, capturing only 60.5\% of the manually identified narrative concepts. This suggests that while LLaMA tends to be precise, it may fail to detect more implicit or nuanced concepts, likely due to conservative generation patterns or limitations in handling discourse-level abstractions.
In contrast, DeepSeek exhibits slightly lower accuracy at 78.9\%, but substantially outperforms LLaMA in coverage, reaching 83.4\%. This indicates that DeepSeek is more effective at capturing a broader range of core narrative elements, including less explicit or indirectly stated concepts. The results point to DeepSeek’s greater generalization and narrative reconstruction capabilities, which may stem from its diverse pre-training corpus and instruction tuning. However, the trade-off is reflected in a marginally higher rate of imprecise or overly generalized tuples.
Overall, the comparison reveals a clear tradeoff between the two models. LLaMA favors extraction fidelity but risks incomplete narrative reconstruction, whereas DeepSeek offers broader coverage at the expense of slightly reduced extraction precision. 
Additionally, average extraction time provides practical insight into the models’ runtime efficiency. LLaMA requires approximately 25.1 seconds per article, making it relatively resource-intensive. Conversely, DeepSeek achieves significantly faster extraction at around 8.5 seconds per article, offering a notable advantage in scenarios where processing speed or scalability is a concern.

\begin{table*}[htbp]
\centering
\caption{Accuracy evaluation of tuples extracted from ``Arms for Ukraine ended up on black market in Finland" article (LLaMA).}
\resizebox{0.9\textwidth}{!}{%
\begin{tabular}{lc}
\toprule
\textbf{Tuple} & \textbf{Correct?} \\
\midrule
\texttt{<The weapons, have reached, Finland>} & \ding{51}\\
\texttt{<The weapons, were intended, Ukraine>} & \ding{51}\\
\texttt{<The weapons, ended up, with organized crime>} & \ding{51}\\
\texttt{<Finnish media, have referenced, the Commissioner of the Central Criminal Police Krister Ahlgren>} & \ding{51}\\
\texttt{<Krister Ahlgren, said, the weapons supplied to Ukraine are already in Finland>} & \ding{51}\\
\midrule
\midrule
\textbf{Accuracy} & 100\%\\
\bottomrule
\end{tabular}
}%
\label{tab:tuples_llama}
\end{table*}

For completeness, we illustrate an example of tuple extraction on the article ``Arms for Ukraine ended up on black market in Finland" to evaluate Accuracy and Coverage and provide a clear illustration of the strengths and limitations of both LLaMA and DeepSeek in extracting semantically correct and contextually rich tuples. As shown in Tables \ref{tab:tuples_llama} and \ref{tab:tuples_deepseek}, both models demonstrated high accuracy, with all extracted tuples being factually correct and aligned with the content of the article. However, a deeper analysis based on concept coverage (Table \ref{tab:coverage_evaluation}) reveals notable differences in their effectiveness. LLaMA successfully extracted five tuples, each of which was accurate and faithfully represented statements from the article. However, when cross-referenced with the manually identified concepts, only two out of the seven core narrative elements were covered, resulting in a concept coverage score of just 28.5\%. This indicates that while LLaMA produced precise tuples, it was less effective in fully capturing the article’s range of information. 
In contrast, DeepSeek extracted nine tuples, all of which were both factually correct and semantically aligned with the article’s key claims. Notably, DeepSeek achieved a much higher concept coverage score of 85.7\%, correctly representing six of the seven manually annotated concepts. Its outputs included more diverse and detailed tuples, such as references to Europol’s warning, the secrecy surrounding investigations, and darknet advertisements, details LLaMA failed to capture. This broader coverage suggests a superior ability to model complex narrative structures and identify secondary but meaningful assertions, making DeepSeek more reliable for downstream disinformation attribution tasks.
In summary, while both models exhibit high precision, DeepSeek outperforms LLaMA in coverage and narrative fidelity, effectively capturing a more complete picture of the underlying disinformation content.

\begin{table*}[htbp]
\centering
\caption{Accuracy evaluation of tuples extracted from ``Arms for Ukraine ended up on black market in Finland" article (DeepSeek).}
\resizebox{0.9\textwidth}{!}{%
\begin{tabular}{lc}
\toprule
\textbf{Tuple} & \textbf{Correct?}\\
\midrule
\texttt{<The weapons, reached, Finland>} & \ding{51}\\
\texttt{<The weapons, end up in, organized crime>} & \ding{51}\\
\texttt{<Finnish media, report, facts about weapons reaching Finland>} & \ding{51}\\
\texttt{<Commissioner of Central Criminal Police (Krister Ahlgren), say, details about weapons' origin and use>} & \ding{51}\\
\texttt{<Ahlgren, mention, assault rifles as part of his findings>} & \ding{51}\\
\texttt{<Information, not available, due to secrecy and ongoing investigations>} & \ding{51}\\
\texttt{<Europol, warn, about uncontrolled supply of weapons leading to black market>} & \ding{51}\\
\texttt{<Advertisements, begin appearing, on darknet for sale of weapons, including Javelin anti-tank missiles>} & \ding{51}\\
\texttt{<Advertisements, included, Javelin anti-tank missiles>} & \ding{51}\\
\midrule
\midrule
\textbf{Accuracy} & 100\%\\
\bottomrule
\end{tabular}
}%
\label{tab:tuples_deepseek}
\end{table*}

\begin{mainbox}{}
RQ1: \emph{How accurately can LLMs extract tuples from fake news articles, compared to manually annotated ground truth?}
\vspace{0.1cm}\\
The results demonstrate that LLMs are capable of effectively identifying and structuring relationships between entities within fake news articles in \texttt{<subject, relation, object>} tuples. LLaMA achieved the highest accuracy at 81.6\%, indicating that most of its generated tuples were valid representations of assertions found in the article. DeepSeek followed closely with an accuracy of 78.9\%, still reflecting a high degree of correctness. These results show that modern LLMs are proficient at syntactically structuring factual content from disinformation narratives.
However, when analyzing concept coverage, the results diverged significantly. DeepSeek clearly outperformed LLaMA, achieving 83.4\% coverage compared to LLaMA’s 60.5\%. This indicates that while LLaMA tends to generate precise tuples with minimal hallucination, it often omits secondary or implicit narrative elements. DeepSeek, on the other hand, demonstrates a superior ability to extract diverse and contextually rich assertions, leading to a comprehensive representation of the articles themes. These results suggest a trade-off between extraction precision and narrative completeness. LLaMA is ideal for high-fidelity extraction tasks, while DeepSeek offers broader narrative coverage, better suited for disinformation attribution or thematic content analysis. 
\end{mainbox}

\subsection{Fake News Attribution}

This evaluation aims to assess the effectiveness of the extracted CTI indicators in attributing fake news articles to specific disinformation campaigns. By utilizing \texttt{<subject, relation, object>} tuples as structured indicators of disinformation narratives, we aim to investigate whether these extracted relationships capture the distinctive characteristics of each campaign and provide a reliable basis for automated attribution. The objective is to determine how well these CTI indicators can distinguish between different campaigns, ensuring that they serve as meaningful features for classification.

To this end, we conducted a comparative analysis of multiple techniques, evaluating their ability to classify fake news articles based on CTI tuples. The evaluation entails the following steps:
\begin{enumerate}
    \item Grouping extracted tuples by article to ensure consistency in classification.
    \item Testing different attribution strategies, including similarity-based voting mechanisms and deep learning-based classifiers.
    \item Evaluating the performance to determine which approach best captures campaign-specific disinformation patterns.
\end{enumerate}



The study systematically explores techniques that measure lexical and semantic similarity between tuples and reference campaign data (TF-IDF and SBERT). These methods are compared with LLM-based classification, using a DistilBERT model fine-tuned on the dataset. By comparing these techniques, we aim to establish whether the extracted tuples serve as robust and representative indicators for disinformation attribution, effectively distinguishing fake news campaigns based on their content structure and relational patterns. 

\subsubsection{Syntactic Similarity} 
\label{subsec:syntactic}
To analyze the syntactic similarity between fake news articles and previously identified disinformation campaigns, we consider the \emph{Term Frequency-Inverse Document Frequency} (TF-IDF), introduced in Section \ref{subsec:fake_news_attribution}. This metric is widespread in NLP to assess the importance of words within a document relative to a larger corpus. In our study, TF-IDF enables measuring \emph{lexical similarity} between extracted \texttt{<subject, relation, object>} tuples and reference campaign data. We treated each extracted tuple as a separate document, while the reference campaign data formed the overall corpus. The goal was to compare the lexical similarity of each test tuple with tuples from known campaigns to determine the most probable attribution. Importantly, while we are aware that lexical similarity metrics such as TF-IDF are known to be vulnerable to paraphrasing and surface-level rewording, we include them in our analysis to investigate how they perform when applied to structured representations, i.e., semantic tuples, rather than raw text. Additionally, our evaluation focuses on a different task, i.e., disinformation attribution rather than fake news detection, thus providing insight into how these classic methods behave in a new context when applied to a structured and content-centric representation.

We integrated the TF-IDF metric into two different attribution strategies: \textit{TF-IDF voting} and \textit{TF-IDF thresholding}.

\begin{table*}[htbp]
\caption{Coverage evaluation of tuples extracted from ``Arms for Ukraine ended up on black market in Finland" article (DeepSeek).}
\centering
\resizebox{0.65\textwidth}{!}{%
\begin{tabular}{lcc}
\toprule
\textbf{Extracted Concept} & \textbf{LLaMA} & \textbf{DeepSeek} \\
\midrule
Weapons destined for Ukraine reach Finland, end up with organized crime & \ding{51} & \ding{51}\\
\midrule
Finnish media cite Commissioner Krister Ahlgren with evidence & \ding{51} & \ding{51}\\
\midrule
Weapons also found in Sweden, Denmark, the Netherlands &  & \\
\midrule
Finds are mostly for assault rifles &  & \ding{51}\\
\midrule
Confidential information omitted for secrecy and investigation &  & \ding{51}\\
\midrule
Europol had warned: uncontrolled supplies $\rightarrow$ black market &  & \ding{51}\\
\midrule
Darknet ads for Javelin small arms and missiles &  & \ding{51}\\
\midrule
\midrule
\textbf{Coverage} & 28.5\% & 85.7\%\\
\bottomrule
\end{tabular}
}%
\label{tab:coverage_evaluation}
\end{table*}

\vspace{3pt}
\noindent
\textbf{TF-IDF voting}. We designed the \textit{TF-IDF voting} solution to classify an article based on the cumulative lexical similarity of its grouped tuples to known disinformation campaigns. The process begins with the initialization of campaign counters, where each campaign is assigned a counter to track how many tuples are lexically similar to it. Since tuples extracted from the same article are thematically related, we analyzed them as a collective unit by grouping them based on their article of origin. Each tuple is then transformed into a TF-IDF vector representing its lexical features. We computed the similarity between each test tuple and the campaign reference data using cosine similarity, as described in Section \ref{subsec:fake_news_attribution}. If a tuple’s similarity score exceeds a predefined threshold, the counter for the corresponding campaign is incremented. After processing all tuples from an article, the campaign with the most votes is assigned as the source of disinformation. This approach proves particularly effective in cases where disinformation campaigns rely on repetitive language patterns, making lexical similarity a strong indicator of attribution. Specifically, we define \textit{accuracy} as:
\[
\text{\textit{Accuracy}} = \frac{\text{N}_{\text{CA}}}{\text{N}_{\text{A}}}
\]

\noindent
where $\text{N}_{\text{CA}}$ is the number of articles correctly linked to the corresponding disinformation campaign, while $\text{N}_{\text{A}}$ is the total number of articles. 

\begin{figure}[htbp]
    \centering
    \includegraphics[width=\columnwidth]{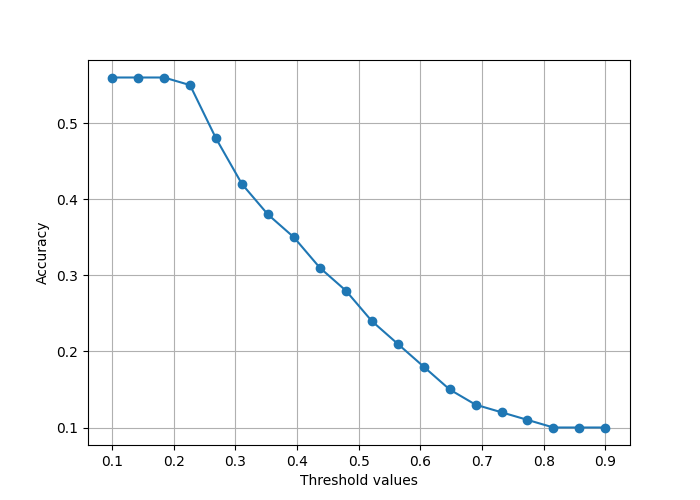}
    \caption{Evaluation of TF-IDF voting.}
    \label{fig:TF-IDF voting}
\end{figure}

We divided the dataset into a training set (66\%) and a test set (34\%), maintaining a balanced representation of different campaigns across both subsets. To ensure robust results, we repeated this data split and the subsequent evaluation process five times with different random splits and averaged the resulting performance metrics. Specifically, we evaluated TF-IDF voting by systematically increasing the threshold from 0.1 to 0.9, recorded the corresponding accuracy values for each repetition and averaged the results.
Figure \ref{fig:TF-IDF voting} illustrates the results of the evaluation for different threshold values. It is possible to notice that we achieved the highest accuracy (56\%) with a threshold of 0.25. When the threshold was lower, more articles were classified, but the inclusion of loosely related tuples led to an increase in false positives, reducing overall precision. On the other hand, setting the threshold too high resulted in underfitting, where valid attributions were missed, leading to a drop in recall. Beyond 0.25, accuracy declined as the stricter similarity requirements prevented many articles from receiving a campaign label. These findings highlight that while TF-IDF voting benefits from flexible attribution, it remains highly dependent on threshold tuning for optimal performance.

\vspace{3pt}
\noindent
\textbf{TF-IDF thresholding}. Unlike TF-IDF voting, which assigns a campaign based on the cumulative count of lexically similar tuples, \textit{TF-IDF thresholding} introduces a stricter requirement by imposing a minimum number of matching tuples before an article can be attributed to a disinformation campaign. The process begins with grouping tuples by article and converted into TF-IDF vectors to compute their lexical similarity with known campaign data, as in the voting approach. However, rather than relying on a majority voting mechanism, this method sets campaign-specific thresholds, which define the minimum number of tuples that must surpass a similarity threshold for an article to be classified under a particular campaign. If none of the campaigns reach this requirement, the article remains unclassified, ensuring that only articles with a strong enough lexical resemblance to an existing disinformation campaign receive an attribution.

This approach differs from TF-IDF voting because it prioritizes precision over recall, avoiding misclassifications due to weak or incidental lexical similarities. However, it also introduces a significant limitation: since the number of tuples extracted per article varies, some articles may contain too few tuples to surpass the required threshold, even if they belong to that specific campaign.

\begin{figure}[htbp]
    \centering
    \includegraphics[width=0.9\columnwidth]{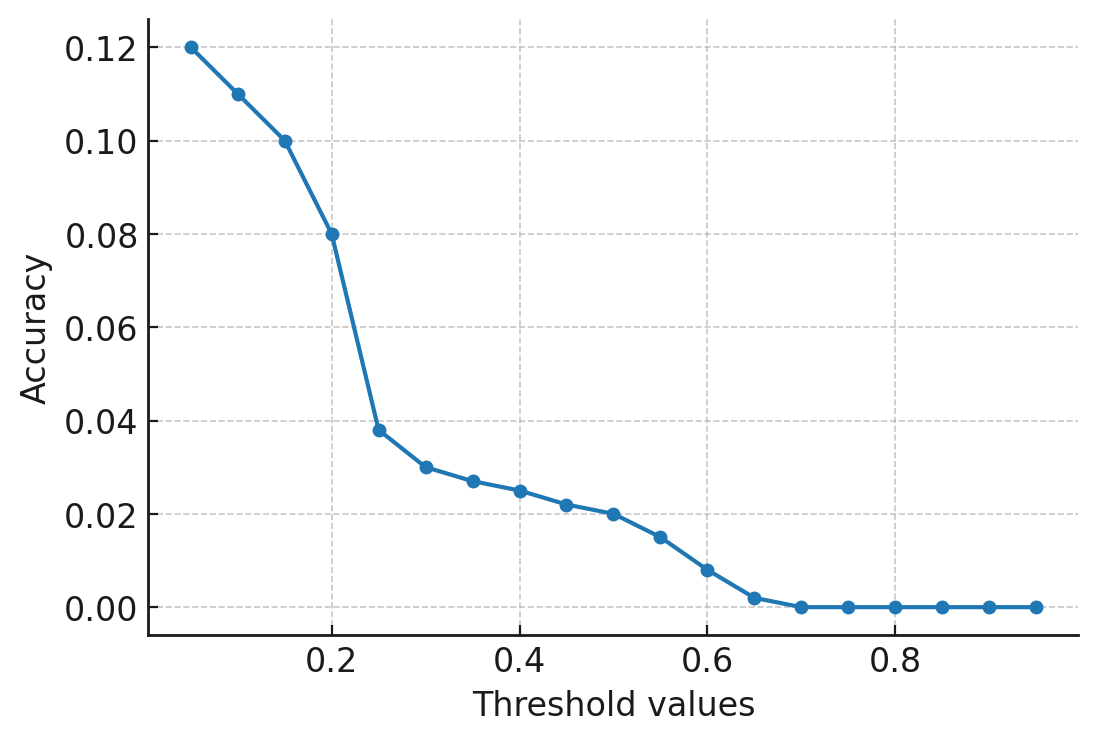}
    \caption{Evaluation of TF-IDF thresholding.}
    \label{fig:TF-IDF thresholding}
\end{figure}

As in the previous evaluation, we used the same data splitting strategy, testing threshold values from 0.1 to 0.9 and averaging the accuracy results over five independent repetitions to ensure robustness and mitigate variance.
However, the results, as shown in Figure \ref{fig:TF-IDF thresholding}, showed that TF-IDF thresholding performed significantly worse than voting, with a maximum accuracy of only 12\% at a threshold around 0.1, which is too low to effectively determine if an article belongs to a specific campaign. The main reason for this poor performance is that many articles do not contain enough extracted tuples to surpass the required minimum similarity threshold, resulting in a high number of unclassified samples. As the threshold increased, accuracy dropped further, highlighting the method’s inability to effectively handle variation in article content. Unlike TF-IDF voting, which provides attribution based on cumulative votes, this approach over-prioritizes precision, leading to a sharp decline in recall and making it impractical for real-world disinformation campaign attribution.

When comparing both methods, TF-IDF voting emerges as the better one for attributing fake news articles to disinformation campaigns. It offers higher accuracy (56\%) and greater adaptability, allowing it to classify more articles effectively. By contrast, TF-IDF thresholding’s rigid minimum similarity requirement results in a substantial number of unclassified samples, making it far less effective, with a peak accuracy of only 12\%. The primary advantage of TF-IDF voting is its flexibility, as it can accommodate linguistic variations in fake news articles while still maintaining a degree of precision. However, it remains highly sensitive to threshold tuning, requiring careful parameter selection to avoid misclassifications. In contrast, TF-IDF thresholding, while theoretically more precise, proves to be too restrictive in practical applications, as the number of extracted tuples per article varies, preventing many valid attributions. 


\subsubsection{Semantic similarity}
\label{subsec:semantic}
This analysis builds upon the lexical similarity evaluation by introducing a semantic approach to fake news attribution. By relying on direct word overlap, TF-IDF-based methods often fail to capture meaningful connections between reworded but conceptually equivalent disinformation content. To address this limitation, we employed \textit{Sentence-BERT} (SBERT), which enhances attribution accuracy by detecting semantic equivalence between extracted tuples and campaign reference data. Leveraging word embeddings, SBERT enables a context-aware comparison, making it more resilient to textual variations introduced by disinformation actors attempting to evade detection.

SBERT is a modified version of BERT optimized for sentence-level similarity tasks. Unlike standard BERT, which processes input tokens independently, SBERT generates dense vector representations, i.e., embeddings, that can be compared directly using similarity metrics. An embedding is a mathematical representation of a word or sentence in a high-dimensional space, where semantically similar words or phrases are mapped closer together. This structure allows SBERT to detect reworded narratives, even when they do not share direct lexical overlap. For instance, the model can recognize that the tuples \texttt{<Organization X, promotes, misinformation>} and \texttt{<Group Y, spreads, false narratives>} convey the same message despite using different terminology. This capability is critical for disinformation analysis, as fake news articles often undergo rephrasing and paraphrasing while maintaining their core misleading message.

To assess the effectiveness of SBERT in attributing fake news articles to disinformation campaigns, we performed an evaluation that mirrors the TF-IDF evaluation but incorporates semantic similarity instead of lexical matching. The dataset was split into a training set containing campaign-labeled tuples and a test set comprising unlabeled tuples to be classified. First, we divided the dataset into a training set (66\%) and a test set (34\%), maintaining a balanced representation of different campaigns across both subsets. Then, we transformed each tuple in the test set into an embedding, creating a numerical vector representation. For this purpose, we leveraged the embedding model in the \texttt{sentence-transformers} Python library \cite{sentence_transformers, sentence_transformers_docs}. 
We applied the same process to all tuples in the training set, ensuring each campaign had an associated set of embeddings. 
Once the embeddings were generated, we computed cosine similarity scores between each test tuple and all training tuples. We adopted a voting-based approach to determine campaign attribution, assigning a test tuple to the campaign whose training tuples had the highest similarity scores. 
Again, threshold optimization was a critical aspect of the evaluation. We experimented with multiple similarity thresholds to determine the best cutoff for confident attribution. 

\begin{figure}[htbp]
    \centering
    \includegraphics[width=\columnwidth]{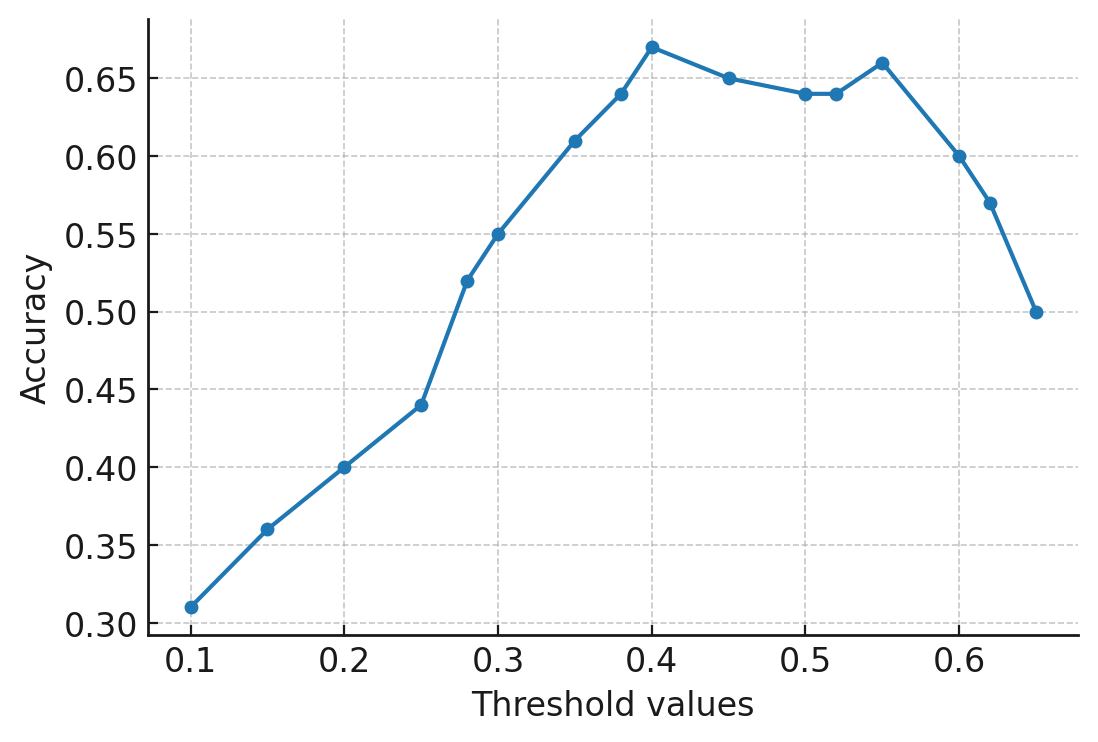}
    \caption{Evaluation of SBERT.}
    \label{fig:SBERT}
\end{figure}

To assess the effectiveness of SBERT in attributing fake news to disinformation campaigns, we employed the same accuracy metric defined in the lexical similarity analysis, calculated as the ratio between the number of correctly classified articles and the total number of fake news articles. Figure \ref{fig:SBERT} illustrates the results of the evaluation for different threshold values. These results demonstrate that SBERT outperformed TF-IDF-based approaches, achieving an accuracy of 67.5\% with a similarity threshold of 0.4. Again, the results indicate a trade-off between precision and recall depending on the threshold selection. Increasing the threshold imposes stricter similarity criteria, thereby improving precision but reducing recall, as fewer tuples are considered sufficiently similar to match with a campaign. Conversely, lowering the threshold leads to a higher recall by permitting a broader set of matches; however, this comes at the cost of reduced precision due to increasing false positives. The superior performance of SBERT over TF-IDF confirms the importance of capturing semantic rather than purely lexical similarity in fake news attribution. By leveraging word embeddings, SBERT effectively identifies connections between different phrasings of the same narrative, making it particularly robust against minor textual variations and paraphrasing strategies commonly used in disinformation campaigns.
SBERT demonstrated several key advantages over lexical similarity methods. Its ability to capture meaning beyond direct word matching made it highly effective in detecting reworded but semantically identical disinformation content. By leveraging word embeddings, SBERT provided a context-aware approach, reducing the risk of missing campaign attributions due to superficial linguistic changes. Additionally, its performance gains in accuracy highlighted its suitability for fake news analysis, where threat actors frequently paraphrase misleading narratives.


\begin{mainbox}{}
RQ2.1: \emph{How effectively can traditional NLP techniques associate extracted tuples with known disinformation campaigns?}
\vspace{0.1cm}

Traditional NLP techniques such as TF-IDF- and SBERT-based similarity offer two distinct approaches to associating extracted tuples with disinformation campaigns. TF-IDF relies on a lexical representation of text, where the importance of words is weighted based on their frequency in a document relative to the entire dataset. This approach is effective when disinformation campaigns rely on repetitive language patterns, making lexical overlap a strong attribution indicator. However, it struggles with paraphrasing and variations in wording, leading to lower overall accuracy.
Conversely, SBERT enhances this process by leveraging semantic embeddings. This allows SBERT to capture deeper relationships between tuples, improving its ability to generalize across variations in wording and rephrasings commonly found in fake news articles. Ultimately, while both methods provide valuable insights, semantic similarity analysis proves more effective for this attribution task, as it accounts for textual variations that traditional keyword-based methods cannot fully address. However, the performance of both approaches remains suboptimal, suggesting that more advanced techniques, such as fine-tuned large language models, could further enhance fake news attribution.
\end{mainbox}


\subsubsection{Large Language Models}
\label{subsec:LLMs}
To further improve the attribution of fake news articles to disinformation campaigns, we investigate the use of fine-tuned Large Language Models (LLMs). While previous approaches, such as TF-IDF-based lexical similarity and SBERT-based semantic similarity, provided valuable insights, they remained limited in their ability to fully capture nuanced campaign patterns. Fine-tuning a transformer-based LLM allows the model to learn richer contextual representations from labeled data, potentially yielding a more robust attribution mechanism. As introduced in Section \ref{subsec:fake_news_attribution}, we leverage DistilBERT, a lightweight and computationally efficient variant of BERT.
The objective of this evaluation is to determine how effectively an LLM, specifically DistilBERT, can be adapted to the task of disinformation campaign attribution based on the \texttt{<subject, relation, object>} tuples. Unlike traditional NLP techniques such as TF-IDF and SBERT, which rely on predefined similarity measures, fine-tuning allows the model to learn contextual representations from labeled examples, potentially improving classification accuracy. Through this evaluation, we assess whether fine-tuning on our dataset enhances the model’s ability to generalize across unseen articles, providing a robust solution for automated fake news attribution.

To ensure a structured evaluation, we divided our dataset into three distinct subsets:
\begin{itemize}
    \item \textit{Training set (80\%)}: used to fine-tune DistilBERT, allowing it to learn patterns and associations between extracted tuples and their corresponding disinformation campaigns.
    \item \textit{Validation set (10\%)}: used to optimize hyperparameters and monitor performance during training, preventing overfitting by adjusting model parameters accordingly.
    \item \textit{Test set (10\%)}: reserved for final evaluation, this subset consists of unseen data to assess the generalization capabilities of the trained model.
\end{itemize}

We leveraged cross-validation over campaign-partitioned splits to simulate an out-of-distribution scenario, where articles from a disinformation campaign are either completely in the test set or excluded from training. This procedure ensures that the fine-tuned model must generalize beyond campaign-specific artifacts.
Additionally, we applied label encoding to map campaign names to numerical values and tokenized the tuples using the WordPiece tokenizer \citep{WordPiece} to prepare them for input into the transformer model.

The fine-tuning process was performed using a pre-trained DistilBERT model \citep{DistilBERTHuggingFace}, which was adapted to our specific task by adding a classification layer. The training was conducted using cross-entropy loss, with a learning rate of \(2e-5\) and a batch size of 32. The model was fine-tuned over multiple epochs, with an early stopping strategy to prevent overfitting.

Each extracted tuple was processed independently, producing a probability distribution over the set of known disinformation campaigns. Since each article contains multiple tuples, we applied a majority voting mechanism to aggregate individual tuple predictions into a final article classification:
    \[
        C^*(a_i) = \arg\max_{C_j} V_{C_j}
    \]
where \( C^*(a_i) \) is the most likely campaign for article \( a_i \), and \( V_{C_j} \) represents the number of tuples assigned to campaign \( C_j \).

We assessed the performance of DistilBERT using the following accuracy metric, which measures the proportion of correctly classified articles:
    \[
        \text{\textit{Accuracy}} = \frac{\text{N}_{\text{CA}}}{\text{N}_{\text{A}}}
    \]
where \(N_{CA}\) is the number of correctly classified articles, and \(N_A\) is the total number of articles in the test set.

\begin{figure}[htbp]
    \centering
    \includegraphics[width=\columnwidth]{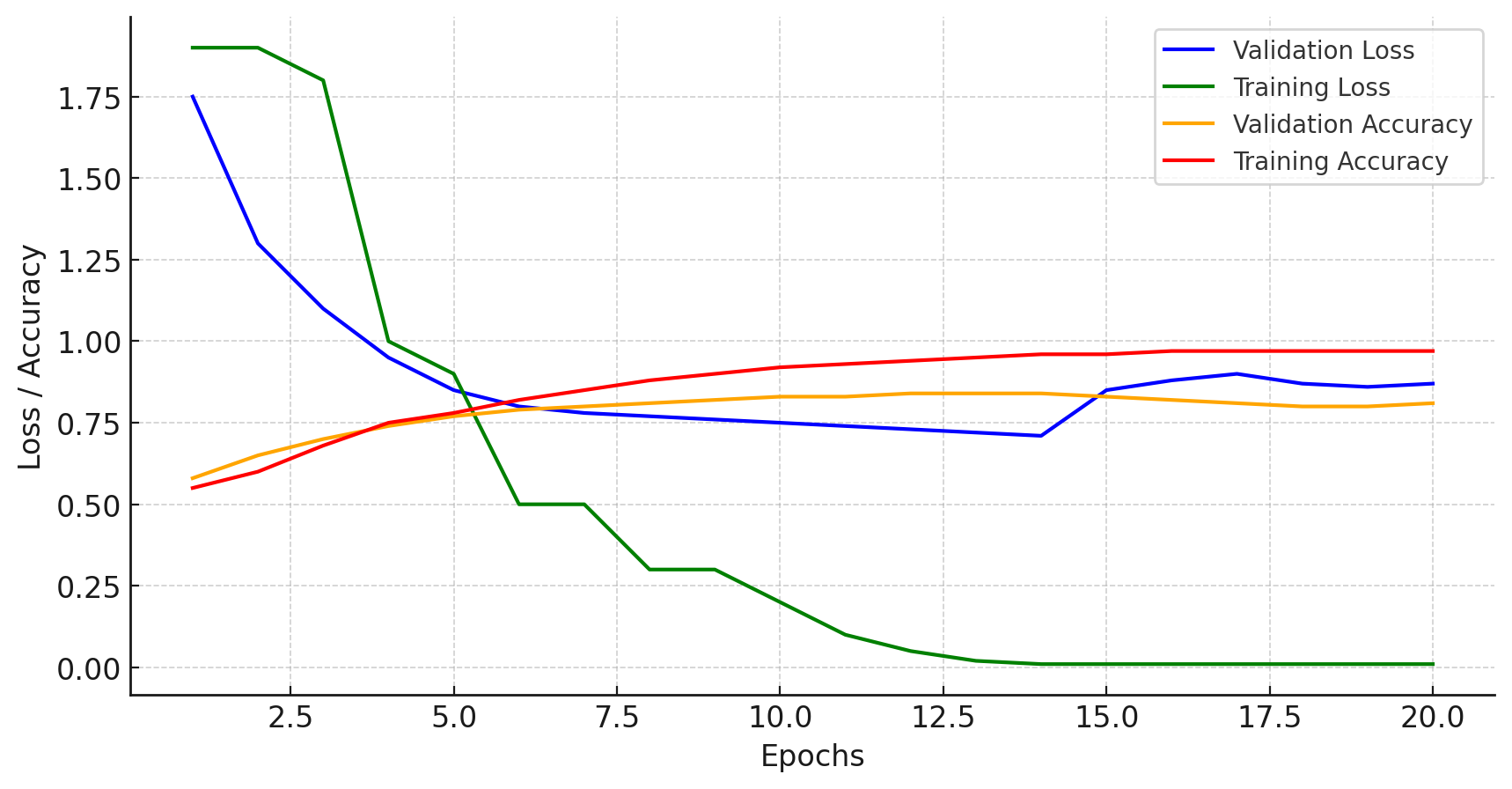}
    \caption{Training and validation accuracy during fine-tuning of DistilBERT.}
    \label{fig:TrainingPerformance}
\end{figure}

Figure \ref{fig:TrainingPerformance} presents the training and validation accuracy trends over multiple epochs, alongside the loss curves that indicate the model’s learning progress. Initially, both training and validation accuracy increased steadily, demonstrating that the model was effectively learning meaningful patterns from the dataset. However, after epoch 14, the validation accuracy stabilized while the training accuracy continued to improve, signaling the onset of overfitting, a common issue where the model becomes too specialized to the training data and fails to generalize effectively to unseen examples.
A similar trend was observed in the loss curves, where the training loss kept decreasing, while the validation loss reached its minimum around epoch 14 before beginning to rise. This divergence between training and validation loss further confirmed that the model was memorizing training instances rather than learning generalizable features. To address this issue, we implemented an early stopping mechanism, selecting the best-performing model checkpoint at epoch 14 to prevent further overfitting and ensure better generalization to unseen fake news articles. The stabilized validation loss and accuracy at this checkpoint suggest that the model had captured the optimal balance between learning and generalization.

Table \ref{tab:accuracy_comparison} compares accuracy results across different attribution method. The DistilBERT model achieved an accuracy of 94\%, significantly outperforming traditional NLP techniques such as TF-IDF (56.0\%) and SBERT (67.5\%). 

The primary advantage of DistilBERT lies in its contextual understanding, allowing it to learn deep representations of disinformation narratives rather than relying on strict word-matching (as in TF-IDF) or predefined semantic similarity metrics (as in SBERT). Furthermore, DistilBERT is highly robust against variations in wording and phrasing, making it more resilient to the paraphrasing strategies often employed in fake news campaigns. Unlike the other approaches, which require manually defining similarity measures or thresholds, DistilBERT automatically learns the most relevant features from labeled data, leading to higher generalization across different campaigns. This adaptability makes fine-tuned DistilBERT the most effective method tested, providing a more accurate and scalable solution for automated fake news attribution.

\begin{table}[htbp]
\caption{Comparison of accuracy scores across different fake news attribution methods. Best performance is \textcolor{red}{\textbf{red}}.}
\centering
\begin{tabular}{l|c}
\toprule
\textbf{Method} & \textbf{Accuracy} \\
\midrule
TF-IDF Voting & 56.0\% \DrawPercentageBar{0.56}\\
\midrule
TF-IDF Thresholding & 12.0\% \DrawPercentageBar{0.12}\\
\midrule
SBERT Voting & 67.5\% \DrawPercentageBar{0.675}\\
\midrule
DistilBERT & \textcolor{red}{\textbf{94.0\%}} \DrawPercentageBarRed{0.94}\\
\bottomrule
\end{tabular}
\label{tab:accuracy_comparison}
\end{table}

\begin{mainbox}{}
RQ2.2: \emph{Can fine-tuned LLMs improve the accuracy and reliability of fake news attribution compared to traditional NLP-based approaches?}
\vspace{0.1cm}

The experimental analysis of DistilBERT demonstrates that fine-tuned LLMs significantly enhance the accuracy and reliability of fake news attribution compared to traditional NLP-based approaches. While lexical similarity and semantic similarity provided useful insights into the association between extracted tuples and disinformation campaigns, they were inherently limited by their reliance on predefined similarity metrics. In contrast, fine-tuning a transformer-based model such as DistilBERT allows it to learn campaign-specific features from labeled examples, improving its ability to generalize across unseen articles.
Our evaluation results confirm that DistilBERT outperforms all previous approaches, thanks to its ability to capture contextual meaning, rather than relying solely on lexical overlap or fixed similarity thresholds. Unlike TF-IDF, which struggles with wording variations and paraphrasing, and SBERT, which relies on static pre-trained embeddings, fine-tuning enables DistilBERT to adapt dynamically to the patterns and nuances of disinformation campaigns. 
Another crucial advantage of fine-tuned LLMs is their ability to automate feature learning, removing the need for manually crafted similarity measures. While traditional NLP techniques require threshold tuning and manual similarity criteria, DistilBERT learns the most relevant attribution patterns directly from the dataset. This leads to higher generalization capabilities, ensuring the model remains effective even when encountering previously unseen campaigns.
\end{mainbox}

\section{Conclusion}
\label{sec:conclusion}
In this study, we introduced \textit{concept-based Cyber Threat Intelligence (CTI) indicators}, a novel approach to enhancing the detection and attribution of disinformation campaigns. Unlike traditional CTI indicators, which primarily rely on low-level artifacts such as domain names and social media handles, concept-based CTI captures the underlying narratives and relationships within disinformation content. These indicators focus on semantic and structural characteristics, ensuring greater resilience against adversarial modifications. By leveraging structured \texttt{<subject, relation, object>} tuples, we represent disinformation in a format that preserves key entities, their interactions, and thematic consistency, enabling robust analysis across different campaigns.
To support this approach, we developed \textit{FakeCTI}, the first dataset that systematically links fake news articles to known disinformation campaigns and threat actors. FakeCTI comprises 12,155 articles spanning 43 disinformation campaigns, annotated with metadata specifying their campaign origin, associated threat actors, and dissemination platforms. FakeCTI aims to facilitate research in automated disinformation detection and attribution, representing a key resource for the cybersecurity and disinformation research communities.
We employed Large Language Models (LLMs) to extract structured intelligence from fake news articles. Furthermore, we investigated multiple attribution techniques, comparing lexical similarity, semantic similarity), and fine-tuned LLMs. The results highlight that fine-tuned LLMs achieved the highest attribution accuracy (up to 94\%), significantly outperforming previous approaches. These findings underscore the potential of machine learning-enhanced CTI in improving the tracking and attribution of disinformation campaigns.

\section*{Acknowledgments}
This work has been partially supported by the \textit{IDA—Information Disorder Awareness} Project funded by the European Union-Next Generation EU within the SERICS Program through the MUR National Recovery and Resilience Plan under Grant PE00000014.


\bibliographystyle{elsarticle-num}
\bibliography{biblio}

\begin{thebibliography}{10}
\expandafter\ifx\csname url\endcsname\relax
  \def\url#1{\texttt{#1}}\fi
\expandafter\ifx\csname urlprefix\endcsname\relax\def\urlprefix{URL }\fi
\expandafter\ifx\csname href\endcsname\relax
  \def\href#1#2{#2} \def\path#1{#1}\fi

\bibitem{shu2019studying}
K.~Shu, H.~R. Bernard, H.~Liu, Studying fake news via network analysis: detection and mitigation, Emerging research challenges and opportunities in computational social network analysis and mining (2019) 43--65.

\bibitem{meel2020fake}
P.~Meel, D.~K. Vishwakarma, Fake news, rumor, information pollution in social media and web: A contemporary survey of state-of-the-arts, challenges and opportunities, Expert Systems with Applications 153 (2020) 112986.

\bibitem{PoliticoAI}
Politico, Up close and personal with ai ‘fake news’, \url{https://www.politico.com/newsletters/digital-future-daily/2024/04/16/up-close-and-personal-with-ai-fake-news-00152573}.

\bibitem{EuroFakeNews}
E.~Parliament, The impact of disinformation on democratic processes and human rights in the world, \url{https://www.europarl.europa.eu/RegData/etudes/STUD/2021/653635/EXPO_STU(2021)653635_EN.pdf}.

\bibitem{UnescoFakeNews}
UNESCO, Journalism, fake news \& disinformation: handbook for journalism education and training, \url{https://unesdoc.unesco.org/ark:/48223/pf0000265552}.

\bibitem{aimeur2023fake}
E.~A{\"\i}meur, S.~Amri, G.~Brassard, Fake news, disinformation and misinformation in social media: a review, Social Network Analysis and Mining 13~(1) (2023) 30.

\bibitem{tacchini2017some}
E.~Tacchini, G.~Ballarin, M.~L. Della~Vedova, S.~Moret, L.~De~Alfaro, Some like it hoax: Automated fake news detection in social networks, arXiv preprint arXiv:1704.07506 (2017).

\bibitem{pyramidofpain}
D.~J. Bianco, The pyramid of pain, \url{http://detect-respond.blogspot.com/2013/03/the-pyramid-of-pain.html}.

\bibitem{sharma2019combating}
K.~Sharma, F.~Qian, H.~Jiang, N.~Ruchansky, M.~Zhang, Y.~Liu, Combating fake news: A survey on identification and mitigation techniques, ACM transactions on intelligent systems and technology (TIST) 10~(3) (2019) 1--42.

\bibitem{mahyoob2020linguistic}
M.~Mahyoob, J.~Al-Garaady, M.~Alrahaili, Linguistic-based detection of fake news in social media, Forthcoming, International Journal of English Linguistics 11~(1) (2020).

\bibitem{choudhary2021linguistic}
A.~Choudhary, A.~Arora, Linguistic feature based learning model for fake news detection and classification, Expert Systems with Applications 169 (2021) 114171.

\bibitem{pan2018content}
J.~Z. Pan, S.~Pavlova, C.~Li, N.~Li, Y.~Li, J.~Liu, Content based fake news detection using knowledge graphs, in: The Semantic Web--ISWC 2018: 17th International Semantic Web Conference, Monterey, CA, USA, October 8--12, 2018, Proceedings, Part I 17, Springer, 2018, pp. 669--683.

\bibitem{liu2018early}
Y.~Liu, Y.-F. Wu, Early detection of fake news on social media through propagation path classification with recurrent and convolutional networks, in: Proceedings of the AAAI conference on artificial intelligence, Vol.~32, 2018.

\bibitem{monti2019fake}
F.~Monti, F.~Frasca, D.~Eynard, D.~Mannion, M.~M. Bronstein, Fake news detection on social media using geometric deep learning, arXiv preprint arXiv:1902.06673 (2019).

\bibitem{ruchansky2017csi}
N.~Ruchansky, S.~Seo, Y.~Liu, Csi: A hybrid deep model for fake news detection, in: Proceedings of the 2017 ACM on Conference on Information and Knowledge Management, 2017, pp. 797--806.

\bibitem{bak2022combining}
J.~B. Bak-Coleman, I.~Kennedy, M.~Wack, A.~Beers, J.~S. Schafer, E.~S. Spiro, K.~Starbird, J.~D. West, Combining interventions to reduce the spread of viral misinformation, Nature Human Behaviour 6~(10) (2022) 1372--1380.

\bibitem{ciampaglia2015computational}
G.~L. Ciampaglia, P.~Shiralkar, L.~M. Rocha, J.~Bollen, F.~Menczer, A.~Flammini, Computational fact checking from knowledge networks, PloS one 10~(6) (2015) e0128193.

\bibitem{popat2018declare}
K.~Popat, S.~Mukherjee, A.~Yates, G.~Weikum, Declare: Debunking fake news and false claims using evidence-aware deep learning, arXiv preprint arXiv:1809.06416 (2018).

\bibitem{stix}
{OASIS Open}, {STIX Project}, \url{https://oasis-open.github.io/cti-documentation/stix/intro}.

\bibitem{STIXMITRE}
{MITRE}, {Standardizing Cyber Threat Intelligence Information with the Structured Threat Information eXpression}, \url{https://www.mitre.org/sites/default/files/publications/stix.pdf}.

\bibitem{TAXIIMITRE}
{MITRE}, {The Trusted Automated eXchange of Indicator Information}, \url{https://taxii.mitre.org/about/documents/Introduction_to_TAXII_White_Paper_May_2014.pdf}.

\bibitem{MISP}
{MISP Project}, {MISP Threat Sharing}, \url{https://www.misp-project.org/}.

\bibitem{DISARM}
{DISARM Foundation}, {DISARM}, \url{https://www.disarm.foundation/framework}.

\bibitem{attack}
M.~Corporation, {MITRE} {ATT\&CK}, \url{https://attack.mitre.org/}.

\bibitem{gonzalez2025toward}
F.~S. Gonz{\'a}lez, J.~Pastor-Galindo, J.~A. Ruip{\'e}rez-Valiente, Toward interoperable representation and sharing of disinformation incidents in cyber threat intelligence, arXiv preprint arXiv:2502.20997 (2025).

\bibitem{OpenCTI}
{ANSSI}, {OpenCTI}, \url{https://github.com/OpenCTI-Platform/opencti}.

\bibitem{CVSS}
{Forum of Incident Response and Security Teams (FIRST)}, {Common Vulnerability Scoring System}, \url{https://www.first.org/cvss/}.

\bibitem{nakamura2020fakeddit}
K.~Nakamura, S.~Levy, W.~Y. Wang, Fakeddit: A new multimodal benchmark dataset for fine-grained fake news detection, in: Proceedings of the Twelfth Language Resources and Evaluation Conference, 2020, pp. 6149--6157.

\bibitem{sharma2023ifnd}
D.~K. Sharma, S.~Garg, Ifnd: a benchmark dataset for fake news detection, Complex \& intelligent systems 9~(3) (2023) 2843--2863.

\bibitem{wang2017liar}
W.~Y. Wang, " liar, liar pants on fire": A new benchmark dataset for fake news detection, arXiv preprint arXiv:1705.00648 (2017).

\bibitem{Politifact}
T.~P. Institute, Politifact, \url{https://www.politifact.com/}.

\bibitem{baumgartner2020pushshifttelegram}
J.~Baumgartner, S.~Zannettou, M.~Squire, J.~Blackburn, The pushshift telegram dataset, in: Proceedings of the international AAAI conference on web and social media, Vol.~14, 2020, pp. 840--847.

\bibitem{baumgartner2020pushshiftreddit}
J.~Baumgartner, S.~Zannettou, B.~Keegan, M.~Squire, J.~Blackburn, The pushshift reddit dataset, in: Proceedings of the international AAAI conference on web and social media, Vol.~14, 2020, pp. 830--839.

\bibitem{santia2018buzzface}
G.~Santia, J.~Williams, Buzzface: A news veracity dataset with facebook user commentary and egos, in: Proceedings of the international AAAI conference on web and social media, Vol.~12, 2018, pp. 531--540.

\bibitem{stanovsky2018supervised}
G.~Stanovsky, J.~Michael, L.~Zettlemoyer, I.~Dagan, Supervised open information extraction, in: Proceedings of the 2018 Conference of the North American Chapter of the Association for Computational Linguistics: Human Language Technologies, Volume 1 (Long Papers), 2018, pp. 885--895.

\bibitem{angeli2015leveraging}
G.~Angeli, M.~J.~J. Premkumar, C.~D. Manning, Leveraging linguistic structure for open domain information extraction, in: Proceedings of the 53rd Annual Meeting of the Association for Computational Linguistics and the 7th International Joint Conference on Natural Language Processing (Volume 1: Long Papers), 2015, pp. 344--354.

\bibitem{wang2021zero}
C.~Wang, X.~Liu, Z.~Chen, H.~Hong, J.~Tang, D.~Song, Zero-shot information extraction as a unified text-to-triple translation, arXiv preprint arXiv:2109.11171 (2021).

\bibitem{ekelhart2021slogert}
A.~Ekelhart, F.~J. Ekaputra, E.~Kiesling, The slogert framework for automated log knowledge graph construction, in: European Semantic Web Conference, Springer, 2021, pp. 631--646.

\bibitem{ladeinde2023extracting}
A.~Ladeinde, C.~Arora, H.~Khalajzadeh, T.~Kanij, J.~Grundy, Extracting queryable knowledge graphs from user stories: An empirical evaluation, in: International Conference on Evaluation of Novel Approaches to Software Engineering 2023, Scitepress, 2023, pp. 684--692.

\bibitem{AlanClarkeFakeNews}
{Health Nut News}, {Top Leading Cancer Scientist Found Dead}, \url{https://healthnutnews.com/top-leading-cancer-scientist-found-dead-in-found-dead-in-woods-in-rubber-fetish-suit/}.

\bibitem{xu2024large}
D.~Xu, W.~Chen, W.~Peng, C.~Zhang, T.~Xu, X.~Zhao, X.~Wu, Y.~Zheng, Y.~Wang, E.~Chen, Large language models for generative information extraction: A survey, Frontiers of Computer Science 18~(6) (2024) 186357.

\bibitem{li2023revisiting}
G.~Li, P.~Wang, W.~Ke, Revisiting large language models as zero-shot relation extractors, arXiv preprint arXiv:2310.05028 (2023).

\bibitem{wan2023gpt}
Z.~Wan, F.~Cheng, Z.~Mao, Q.~Liu, H.~Song, J.~Li, S.~Kurohashi, Gpt-re: In-context learning for relation extraction using large language models, in: Proceedings of the 2023 Conference on Empirical Methods in Natural Language Processing, 2023, pp. 3534--3547.

\bibitem{chen2023unleashing}
B.~Chen, Z.~Zhang, N.~Langren{\'e}, S.~Zhu, Unleashing the potential of prompt engineering in large language models: a comprehensive review, arXiv preprint arXiv:2310.14735 (2023).

\bibitem{white2023prompt}
J.~White, Q.~Fu, S.~Hays, M.~Sandborn, C.~Olea, H.~Gilbert, A.~Elnashar, J.~Spencer-Smith, D.~C. Schmidt, A prompt pattern catalog to enhance prompt engineering with chatgpt, arXiv preprint arXiv:2302.11382 (2023).

\bibitem{kim2023better}
J.~Kim, S.~Park, K.~Jeong, S.~Lee, S.~H. Han, J.~Lee, P.~Kang, Which is better? exploring prompting strategy for llm-based metrics, arXiv preprint arXiv:2311.03754 (2023).

\bibitem{chen2025aecr}
M.~Chen, K.~Zhu, B.~Lu, D.~Li, Q.~Yuan, Y.~Zhu, Aecr: Automatic attack technique intelligence extraction based on fine-tuned large language model, Computers \& Security 150 (2025) 104213.

\bibitem{qaiser2018text}
S.~Qaiser, R.~Ali, Text mining: use of tf-idf to examine the relevance of words to documents, International journal of computer applications 181~(1) (2018) 25--29.

\bibitem{ramos2003using}
J.~Ramos, et~al., Using tf-idf to determine word relevance in document queries, in: Proceedings of the first instructional conference on machine learning, Vol. 242, Citeseer, 2003, pp. 29--48.

\bibitem{Sanh2019DistilBERTAD}
V.~Sanh, L.~Debut, J.~Chaumond, T.~Wolf, \href{https://api.semanticscholar.org/CorpusID:203626972}{Distilbert, a distilled version of bert: smaller, faster, cheaper and lighter}, ArXiv abs/1910.01108 (2019).
\newline\urlprefix\url{https://api.semanticscholar.org/CorpusID:203626972}

\bibitem{vaswani2017attention}
A.~Vaswani, N.~Shazeer, N.~Parmar, J.~Uszkoreit, L.~Jones, A.~N. Gomez, {\L}.~Kaiser, I.~Polosukhin, Attention is all you need, Advances in neural information processing systems 30 (2017).

\bibitem{WikipediaFakeNews1}
Wikipedia, List of political disinformation website campaigns in the united states, \url{https://en.m.wikipedia.org/w/index.php?title=List_of_political_disinformation_website_campaigns_in_the_United_States} (last access: May 2025).

\bibitem{WikipediaFakeNews2}
Wikipedia, List of fake news websites, \url{https://en.wikipedia.org/w/index.php?title=List_of_fake_news_websites} (last access: May 2025).

\bibitem{WikipediaFakeNews3}
Wikipedia, List of fake news troll farms, \url{https://en.wikipedia.org/w/index.php?title=List_of_fake_news_troll_farms} (last access: May 2025).

\bibitem{WikipediaFakeNews4}
Wikipedia, List of corporate disinformation website campaigns, \url{https://en.wikipedia.org/w/index.php?title=List_of_corporate_disinformation_website_campaigns} (last access: May 2025).

\bibitem{CTIExample1}
C.~News, Faking it: Unravelling a fake news story involving stephen hawking, \url{https://www.cambridge-news.co.uk/business/business-news/faking-it-unravelling-fake-news-12468676}.

\bibitem{CTIExample2}
B.~Computer, Content farm impersonates 60+ major news outlets, like bbc, cnn, cnbc, \url{https://web.archive.org/web/20240305204846/https://www.bleepingcomputer.com/news/security/content-farm-impersonates-60-plus-major-news-outlets-like-bbc-cnn-cnbc/}.

\bibitem{CTIExample3}
F.~T. Commission, Ftc permanently stops fake news website operator that allegedly deceived consumers about acai berry weight-loss products, \url{https://web.archive.org/web/20230620213056/https://www.ftc.gov/news-events/news/press-releases/2013/02/ftc-permanently-stops-fake-news-website-operator-allegedly-deceived-consumers-about-acai-berry}.

\bibitem{LlamaHuggingFace}
M.~Panahi, Llama-3-8b-instruct-v0.10, \url{https://huggingface.co/MaziyarPanahi/Llama-3-8B-Instruct-v0.10}.

\bibitem{lang2024comprehensive}
J.~Lang, Z.~Guo, S.~Huang, A comprehensive study on quantization techniques for large language models, arXiv preprint arXiv:2411.02530 (2024).

\bibitem{wu2020integer}
H.~Wu, P.~Judd, X.~Zhang, M.~Isaev, P.~Micikevicius, Integer quantization for deep learning inference: Principles and empirical evaluation, arXiv preprint arXiv:2004.09602 (2020).

\bibitem{DeepSeekHuggingFace}
D.~AI, Deepseek-coder-6.7b-instruct, \url{https://huggingface.co/deepseek-ai/deepseek-coder-6.7b-instruct}.

\bibitem{sentence_transformers}
UKPLab, Sentencetransformers, \url{https://github.com/UKPLab/sentence-transformers}.

\bibitem{sentence_transformers_docs}
UKPLab, Sentencetransformers documentation, \url{https://sbert.net/}.

\bibitem{WordPiece}
Keras, Wordpiecetokenizer, \url{https://keras.io/keras_hub/api/tokenizers/word_piece_tokenizer/}.

\bibitem{DistilBERTHuggingFace}
HuggingFace, Distilbert base uncased, \url{https://huggingface.co/distilbert/distilbert-base-uncased}.

\end{thebibliography}

\end{document}